\documentclass[acmsmall]{acmart}
\usepackage{array}
\usepackage{pifont}
\usepackage{colortbl}
\usepackage{xcolor}
\usepackage{enumitem}
\usepackage{booktabs} 
\usepackage{multirow}
\usepackage{tabularx}
\usepackage{makecell}
\usepackage{tikz}
\usepackage{graphicx}
\usepackage{subcaption}
\usepackage{tikz}
\usepackage[normalem]{ulem}
\usepackage{cleveref}
\usepackage{longtable}
\usepackage{float}
\usepackage{lscape}
\usepackage{graphicx}
\PassOptionsToPackage{margin=1in}{geometry}

\definecolor{oxfordblue}{rgb}{0.0, 0.13, 0.28}
\definecolor{harvardcrimson}{rgb}{0.79, 0.0, 0.09}
\definecolor{dartmouthgreen}{rgb}{0.05, 0.5, 0.06}
\definecolor{princetonorange}{rgb}{1.0, 0.56, 0.0}
\definecolor{yaleblue}{rgb}{0.06, 0.3, 0.57}
\definecolor{usccardinal}{rgb}{0.6, 0.0, 0.0}
\definecolor{uclablue}{rgb}{0.33, 0.41, 0.58}
\definecolor{msugreen}{rgb}{0.09, 0.27, 0.23}
\definecolor{cornellred}{rgb}{0.7, 0.11, 0.11}
\definecolor{pomegranate}{RGB}{192, 57, 43}
\definecolor{anti-pomegranate}{RGB}{43,178,192}
\definecolor{alizarin}{RGB}{231, 76, 60}
\definecolor{peter}{RGB}{52, 152, 219}
\definecolor{green}{RGB}{22, 160, 133}
\definecolor{anti-green}{RGB}{160,22,118}
\definecolor{turquoise}{RGB}{26, 188, 156}
\definecolor{pumpkin}{RGB}{211, 84, 0}
\definecolor{anti-pumpkin}{RGB}{0,22,211}
\definecolor{carrot}{RGB}{230, 126, 34}
\definecolor{wisteria}{RGB}{142, 68, 173}
\definecolor{anti-wisteria}{RGB}{99,173,68}
\definecolor{amethyst}{RGB}{155, 89, 182}
\definecolor{sunflower}{RGB}{241, 196, 15}
\definecolor{MAIMIAOLV}{RGB}{85, 187, 138}
\definecolor{charlieblue}{RGB}{34, 66, 148}
\definecolor{darkblue}{RGB}{0, 43, 186}
\definecolor{lightblue}{RGB}{28, 128, 200}
\definecolor{capriblue}{HTML}{1B579C}

\definecolor{safeorange}{HTML}{fc8d59}
\definecolor{safeblue}{HTML}{91bfdb}
\definecolor{safegreen}{HTML}{a1d76a}

\definecolor{qualitative1}{HTML}{8dd3c7}
\definecolor{qualitative2}{HTML}{ffc300}
\definecolor{qualitative3}{HTML}{bebada}
\definecolor{qualitative4}{HTML}{fb8072}
\definecolor{qualitative5}{HTML}{80b1d3}

\definecolor{appleredlight}{RGB}{255, 105, 97}
\definecolor{appleorangelight}{RGB}{255, 179, 64}
\definecolor{appleyellowlight}{RGB}{255, 212, 38}
\definecolor{applegreenlight}{RGB}{48, 219, 91}
\definecolor{applemintlight}{RGB}{102, 212, 207}
\definecolor{appleteallight}{RGB}{93, 230, 255}
\definecolor{applecyanlight}{RGB}{112, 215, 255}
\definecolor{applebluelight}{RGB}{64, 156, 255}
\definecolor{appleindigolight}{RGB}{125, 122, 255}
\definecolor{applepurplelight}{RGB}{218, 143, 255}
\definecolor{applepinklight}{RGB}{255, 100, 130}
\definecolor{applebrownlight}{RGB}{181, 148, 105}

\definecolor{applerednormal}{RGB}{255, 69, 58}
\definecolor{appleorangenormal}{RGB}{255, 159, 10}
\definecolor{appleyellownormal}{RGB}{255, 214, 10}
\definecolor{applegreennormal}{RGB}{48, 209, 88}
\definecolor{applemintnormal}{RGB}{99, 230, 226}
\definecolor{appletealnormal}{RGB}{64, 200, 224}
\definecolor{applecyannormal}{RGB}{100, 210, 255}
\definecolor{applebluenormal}{RGB}{10, 132, 255}
\definecolor{appleindigonormal}{RGB}{94, 92, 230}
\definecolor{applepurplenormal}{RGB}{191, 90, 242}
\definecolor{applepinknormal}{RGB}{255, 55, 95}
\definecolor{applebrownnormal}{RGB}{172, 142, 104}

\definecolor{applegrey}{RGB}{99, 99, 102}

\newcommand{\yh}[1]{{#1}}

\newcommand{\zyh}[1]{{#1}}

\newcommand{\eg}{{\textit{e.g.}}}

\definecolor{green1}{HTML}{b2e2e2}
\definecolor{green2}{HTML}{66c2a4}
\definecolor{green3}{HTML}{238b45}

\newcommand{\smallscriptsize}{\fontsize{6.5pt}{7.5pt}\selectfont}
\newcommand{\Rule}{\textcolor{green1}{\rule{1.5ex}{1.5ex}}}
\newcommand{\ML}{\textcolor{green2}{\rule{1.5ex}{1.5ex}}}
\newcommand{\GenAIInvolved}{\textcolor{green3}{\rule{1.5ex}{1.5ex}}}
\newcommand{\blank}{\textcolor{white}{\rule{1.5ex}{1.5ex}}}

\newcommand{\PublicFunction}{\textcolor{safeblue}{\rule{1.5ex}{1.5ex}}}
\newcommand{\PrivateFunction}{\textcolor{safeorange}{\rule{1.5ex}{1.5ex}}}

\newcommand{\Training}{\textcolor{qualitative1}{\rule{1.5ex}{1.5ex}}}
\newcommand{\Calibrating}{\textcolor{qualitative2}{\rule{1.5ex}{1.5ex}}}
\newcommand{\Customizing}{\textcolor{qualitative3}{\rule{1.5ex}{1.5ex}}}
\newcommand{\Configuring}{\textcolor{qualitative4}{\rule{1.5ex}{1.5ex}}}

\newcommand{\Observation}{\textcolor{qualitative1}{\rule{1.5ex}{1.5ex}}}
\newcommand{\Questionnaire}{\textcolor{qualitative2}{\rule{1.5ex}{1.5ex}}}
\newcommand{\Interview}{\textcolor{qualitative3}{\rule{1.5ex}{1.5ex}}}

\newcolumntype{C}[1]{>{\centering\arraybackslash}p{#1}}
\newcolumntype{?}{!{\color{applegrey!15}\vrule width 0.3pt}}

\setcopyright{acmlicensed}
\copyrightyear{2018}
\acmYear{2018}
\acmDOI{XXXXXXX.XXXXXXX}

\acmJournal{PACMHCI}
\acmVolume{37}
\acmNumber{4}
\acmArticle{111}
\acmMonth{8}

\begin{document}

\title{Designing AI-Infused Interactive Systems for Online Communities: A Systematic Literature Review}


\author{Yuanhao Zhang}
\affiliation{%
  \institution{The Hong Kong University of Science and Technology}
  \city{Hong Kong}
  \country{Hong Kong}}
\email{yzhangiy@connect.ust.hk}

\author{Xiaoyu Wang}
\affiliation{%
  \institution{The Hong Kong University of Science and Technology}
  \city{Hong Kong}
  \country{Hong Kong}
}

\author{Jiaxiong Hu}
\affiliation{%
 \institution{The Hong Kong University of Science and Technology}
 \city{Hong Kong}
 \country{Hong Kong}}

\author{Ziqi Pan}
\affiliation{%
  \institution{The Hong Kong University of Science and Technology}
  \city{Hong Kong}
  \country{Hong Kong}}

\author{Zhenhui Peng}
\affiliation{%
  \institution{Sun Yat-sen University}
  \city{Zhuhai}
  \state{Guangdong}
  \country{China}}

\author{Xiaojuan Ma}
\affiliation{%
  \institution{The Hong Kong University of Science and Technology}
  \city{Hong Kong}
  \country{Hong Kong}}


\renewcommand{\shortauthors}{Trovato et al.}

\begin{abstract}
  
AI-infused systems have demonstrated remarkable capabilities in addressing diverse human needs within online communities. Their widespread adoption has shaped user experiences and community dynamics at scale. However, designing such systems requires a clear understanding of user needs, careful design decisions, and robust evaluation. In this work, we present a systematic review of 77 studies \zyh{on AI-infused systems in online communities}, analyzing \zyh{them} through three lenses: the challenges they aim to address, their design functionalities, and the evaluation strategies employed. The first two dimensions are organized around four core aspects of community participation: contribution, consumption, mediation, and moderation. \zyh{Our synthesis distills a set of key design lessons and considerations, including the necessity to design for broader community ecosystems and prioritize user emotional experiences. We conclude by outlining future directions, such as newcomer socialization and inclusive design.}

\end{abstract}






\maketitle

\section{Introduction}
\label{chap:introduction}

Since their emergence, online communities have become one of the most prominent and enduring forms of digital engagement globally \cite{malinen2015understanding, chen2011moderated, iriberri2009life}. They offer spaces for social support, relationship building, and entertainment \cite{chiu2015understanding, armstrong2009real}. However, sustaining thriving online communities poses persistent challenges, such as fostering continuous member engagement \cite{malinen2015understanding, iriberri2009life}, ensuring the quality of user-generated content (UGC) \cite{chen2011moderated, malinen2015understanding}, and maintaining healthy social dynamics \cite{benke2020chatbot, ridings2010online, sundaram2012understanding}. Addressing these challenges through human efforts alone often falls short due to limitations in scalability, consistency, and responsiveness \cite{choi2023convex, he2023cura, shin2022exploring, lee2020solutionchat}.
Researchers in Computer-Supported Cooperative Work (CSCW) and Human-Computer Interaction (HCI) have proposed a wide range of interactive systems that harness AI capabilities (referred to as \textit{AI-infused systems} in the scope of this paper) \zyh{to mediate UGC, shape social relationships, and enforce community norms.} 
For instance, AI-infused systems can support content creators in generating text or images \cite{peng2020exploring, zhang2024mentalimager}, assist moderators by streamlining the decision-making process \cite{chandrasekharan2019crossmod, rifat2024combating}, and help content consumers to organize and navigate unstructured UGC \cite{zhang2025coknowledge, liu2023coargue}.
The pervasive deployment and large-scale influence of such systems make their responsible design a matter of critical importance.

However, designing effective AI-infused systems is far from straightforward, as it involves clearly identifying community needs, making informed design decisions, and rigorously evaluating the system’s impact \cite{gregor2013positioning, amershi2019guidelines, hevner2004design}. 
\yh{Recent years have witnessed not only a proliferation of such systems but also their technological evolution and role transformation. Consequently,} these systems often differ widely in their design goals, functionalities, and evaluation strategies. Without a structured synthesis of these studies, \yh{it remains unclear how traditional design wisdom applies to these emerging, complex interactions.}
Thus, we identify the need for a systematic analysis of how AI-infused systems operate within online communities. \yh{By distilling
emerging sociotechnical design patterns, our review aims} to support diverse stakeholders and inform the design of more effective, adaptable, and responsible community technologies.
While existing surveys have investigated various forms of human–AI interaction \cite{shi2023understanding, li2024we}, little is known about how such interaction unfolds within online communities, where complex social roles and diverse interaction patterns fundamentally shape the nature and impact of these systems. 
To fill the void, this paper aims to address the following research questions:
\begin{description}
    \item[\hspace{1.5em}RQ1] \textit{How have AI‑infused systems been designed to address challenges in online communities?}
    \item[\hspace{1.5em}RQ2] \textit{How have existing studies evaluated the impact of AI-infused systems in online communities?}
\end{description}

To answer the questions, we constructed a corpus of 77 studies that propose AI-infused systems for online communities, following the PRISMA framework \cite{tugwell2021prisma}.
We then systematically analyzed this corpus to uncover the design patterns and assessment practices. 
\yh{Based on this synthesis, this paper offers the following contributions:

\begin{itemize}
    \item A comprehensive taxonomy of AI functionalities addressing challenges across four typical community behaviors: contribution, consumption, mediation, and moderation (Table \ref{tab:codebook2}). We map these functionalities to specific user challenges, revealing how AI serves as a multifaceted support for community sustainability.

    \item A structured synthesis of evaluation strategies, categorizing adopted metrics into five dimensions: individual-level, community-level, UGC-centric, system-centric, and system usability \& UX (\Cref{fig: RQ2 overview tree diagram 2col} \& \Cref{appendix: RQ2 evaluation metrics}). This categorization clarifies the link between system goals and assessment methods, aiding researchers in designing rigorous evaluations.

    \item Key design lessons derived from identifying tensions in the design space. We highlight the necessity of: (1) designing for the ecosystem; (2) motivating quality over quantity; (3) balancing prioritization with diversity; (4) preserving interaction authenticity; (5) manageing output via progressive disclosure; and (6) pairing dissent with emotional safeguards.

    \item Actionable design considerations and future research opportunities. We discuss critical factors such as emotion experience and content safety, and identify underexplored frontiers including supporting newcomers, inclusive design for diverse abilities , and ethical evaluation of AI in communities.
\end{itemize}
}

\section{Related Work}

\subsection{The Landscape of Online Communities}

Online communities gained prominence in the 1990s with the rise of the Internet and have since become integral to digital life \cite{preece2000online}. Today, they encompass a broad spectrum of platforms, including Q\&A sites (e.g., Stack Overflow), social media groups (e.g., Facebook Groups), gaming guilds (e.g., World of Warcraft), fan forums, professional networks (e.g., LinkedIn), and collaborative knowledge bases (e.g., Wikipedia), with some communities reaching tens of millions of members \cite{oksanen2024online}. These spaces offer educational resources, peer support, and entertainment—services often difficult to replicate in purely offline settings \cite{vickery2007participative, preece2000online, kraut2012building}. A defining feature of these communities is their reliance on UGC. Unlike traditional media, where content is produced by professionals, these communities are driven by everyday users contributing posts, comments, videos, reviews, and other forms of media \cite{chen2011moderated}. This participatory model democratizes content creation and fuels platform growth \cite{preece2000online, vickery2007participative}. 
However, such openness demands ongoing efforts to maintain member engagement, manage complex social dynamics, and ensure both content quality and quantity —issues intensified by the scale and diversity of community members and content \cite{malinen2015understanding, chen2011moderated, kraut2012building}. 

\yh{In response, the integration of AI in online communities continues to evolve in both technology and role.
Historically, community interventions primarily relied on rule-based systems (e.g., regular expressions and blacklists) to automate tasks. 
Tools like Reddit's AutoModerator \cite{jhaver2019human, jhaver2019does} exemplified this era of opaque background enforcement, removing content based on keywords without nuance.
Later, the field shifted towards probabilistic models and early neural networks~\cite{chandrasekharan2019crossmod}, which improved accuracy but largely remained black boxes with limited human oversight.

As collaborative approaches are identified as being more trustworthy~\cite{bhuiyan2021nudgecred}, we are entering a new era where AI is infused into the community member experience to augment human capabilities. For instance, AI now assists moderators in detecting subtle radicalization patterns~\cite{govers2023down} or supports users in assessing news credibility via crowdsourced nudges (e.g., NudgeCred~\cite{bhuiyan2021nudgecred}).
Beyond functional tools, LLMs are enabling AI to evolve into social actors that provide peer-like emotional support \cite{wang2021cass} or facilitate group discussions \cite{shin2023introbot}.

However, this shift introduces new risks. For example, while AI aims to curb extremism, poorly designed mediation can inadvertently exacerbate polarization~\cite{govers2023down}. Similarly, credibility tools can lead to undue reliance, where users over-trust the system's authority even when it errs~\cite{bhuiyan2021nudgecred}.
Thus, prior design wisdom focused on efficiency is insufficient for this paradigm, where AI actively shapes social dynamics, user agency, and community norms. Our review bridges this gap by distilling emerging sociotechnical design patterns. 
By mapping the trajectory from rule-based governance to human-in-the-loop augmentation, we derive actionable design lessons to ensure this evolution steers toward empowerment rather than disruption.}

\subsection{AI-Infused System Research}

Advances in AI technologies have led HCI researchers to explore AI-infused systems through a growing body of empirical work. Some of these studies aim to better understand human–AI interaction patterns. For example, Parasuraman et al. \cite{parasuraman2000model} proposed a ten-level taxonomy of human-autonomy interaction, ranging from fully manual control by humans to complete automation without human involvement. Similarly, Cimolino and Graham \cite{cimolino2022two} introduced a multidimensional framework to characterize human–AI interaction along four axes: AI Role, Supervision, Influence, and Mediation. In parallel, other researchers have proposed principles, frameworks, and design guidelines to support the design of effective AI-infused systems. Amershi et al. \cite{amershi2019guidelines}, for instance, categorized human–AI interaction into four temporal stages—initial interaction, ongoing use, error situations, and long-term adaptation—and provided corresponding design recommendations for each stage, such as ``remember recent interactions'' and ``notify users about changes.''

While these general, high-level understandings are valuable, domain-specific insights into AI capabilities and roles can offer more actionable guidance for designing systems in particular contexts. For example, Li et al. \cite{li2024we} deconstructed the data storytelling process into four stages—analysis, planning, implementation, and communication—and identified the respective roles of humans and AI in each. Zheng et al. \cite{zheng2022ux} focused specifically on conversational agents, examining their capabilities and the evaluation strategies used to assess their performance. Shi et al. \cite{shi2023understanding} examined how AI assists designers in diverse design tasks, such as visualizing concepts, generating content, and testing outcomes. However, despite this growing literature, there remains a lack of systematic reviews addressing AI-infused systems in the context of online communities—a setting characterized by highly heterogeneous user populations and diverse user engagement behavior. The unique needs, design strategies, and evaluation approaches in this domain likely differ from other settings, warranting closer attention, which this survey aims to provide.

\section{Methodology}

This section outlines our research methodology, including the procedures for paper collection and data analysis.

\subsection{Paper Collection}

\begin{figure*}[htbp]
    \centering
    \includegraphics[width=1\textwidth]{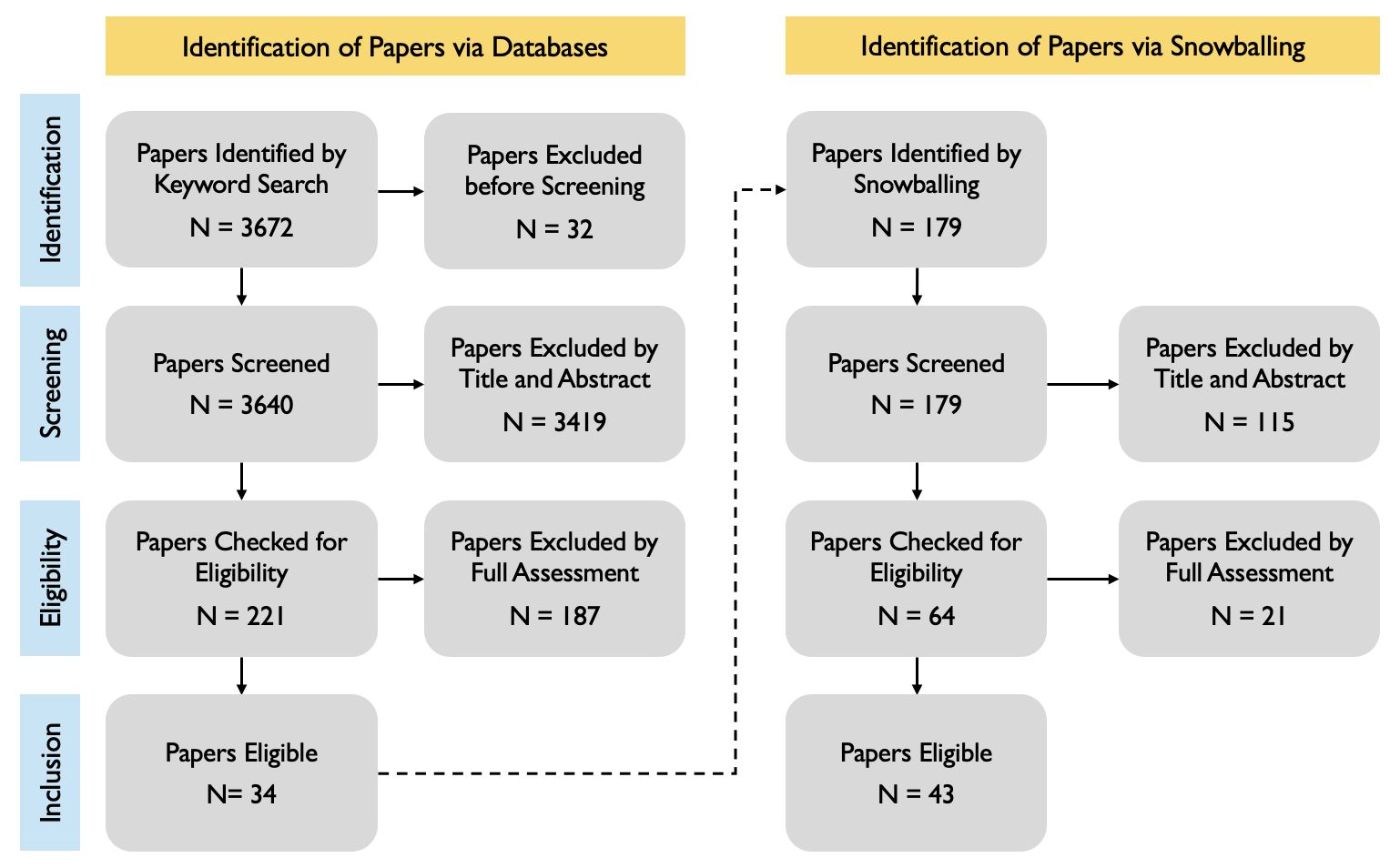}
    \caption{Flowchart of our paper collection process following PRISMA guidelines.}
    \Description{}
    \label{fig: PRISMA flowchart}
\end{figure*}

Following the PRISMA guidelines \cite{tugwell2021prisma} and drawing on established practices from prior work \cite{jalali2012systematic, kitchenham2007guidelines, webster2002analyzing}, we employed a multi-step process to curate a relevant corpus. We began with systematic database searches using predefined search strings to identify relevant studies. To ensure comprehensiveness, we then applied backward snowballing (reviewing reference lists) and forward snowballing (examining citing papers) to supplement the initial results. \Cref{fig: PRISMA flowchart} illustrates the full selection process.

\subsubsection{Identification}
We conducted a systematic search to identify papers on AI-infused systems in online communities, using keyword queries across major databases including ACM Digital Library, IEEE Xplore, Springer, Taylor \& Francis, and ScienceDirect—selected for their strong coverage of HCI, CSCW, visualization, and information systems research, which were particularly relevant to our topic. 
Inspired by prior reviews on human-AI interaction \cite{shi2023understanding, zheng2022ux} and online communities\cite{sun2014understanding, akar2018user}, we crafted the search query and structured it into two components: 
1) The first component targeted common variations and synonyms of \textbf{online communities}: \textit{(``online communit*''} OR \textit{``digital communit*''} OR \textit{``virtual communit*''} OR \textit{``group chat''} OR \textit{``social network''} OR \textit{``social media'')}. 
2) The second component captured keywords related to the \textbf{AI-infused systems}: \textit{(support*} OR \textit{assist*} OR \textit{tool} OR \textit{prototype} OR \textit{artifact} OR \textit{system} OR \textit{``conversational agent''} OR \textit{``intelligent assistant''} OR \textit{``intelligent agent''} OR \textit{bot} OR \textit{``conversational AI''} OR \textit{``human-AI collaboration'')}.
These terms were searched within the title, abstract and author keywords fields. The detailed query used for each database depends on platform-specific policies.
The search, completed in April 2025, yielded a total of 3640 papers.


\subsubsection{Screening}
\label{sec: 3.1.2}

Two authors (henceforth analysts) independently screened the initial set of papers based on titles and abstracts, applying the following inclusion and exclusion criteria: 
1) The papers must be written in English. 
2) The papers must be peer-reviewed and substantive, excluding non-archival works and unreviewed materials. 
3) The study must be situated in the context of online communities. 
We adopt an inclusive definition of online communities as: \textit{``cyberspace[s] supported by computer-based information technology, centered upon communication and interaction of participants to generate member-driven content, resulting in a relationship being built''} \cite{lee2003virtual}.
\yh{This definition emphasizes UGC as the medium of interaction (thereby excluding purely functional communication tools like video conferencing apps) and the emergence of social relationships (excluding transactional platforms like e-commerce or transient task crowdsourcing where no communal bond is formed).
Under these constraints, a broad scope is essential to capture an evolving landscape where platform boundaries blur (e.g., forums vs. chats). It enables us to synthesize AI design patterns that address fundamental challenges related to content and relationships shared across diverse architectures.}
4) The papers should describe an AI-infused system. To align with the focus of this review, we limit our scope to human-centric, AI-supported tools, excluding studies that are purely empirical or do not involve system design.
We define AI broadly as \textit{``systems that display intelligent behavior by analyzing their environment and taking actions - with some degree of autonomy - to achieve specific goals''} \cite{sheikh2023artificial}, where heuristics-based AI, ML-based AI, and GenAI are all included. 
\yh{We adopt this comprehensive definition because a system's impact in online communities often stems from its perceived function and social role rather than algorithmic complexity alone. This inclusive scope ensures the retention of foundational tools that significantly shape community dynamics, capturing the full spectrum of sociotechnical interventions used to address diverse community challenges.}
Uncertainties during screening were resolved through discussion between the analysts. This process resulted in a corpus of 221 papers.

\subsubsection{Eligibility}

To assess eligibility, the two analysts independently conducted full-text reviews of the screened papers and resolved discrepancies through multiple rounds of discussion. During this process, we refined our exclusion criteria based on observed edge cases:
1) We excluded papers that introduced AI-infused systems related to online communities but did not serve community-centered goals—such as systems designed solely to promote offline activities (e.g., Botivist \cite{savage2016botivist}) or to extract data from online communities for unrelated purposes (e.g., \cite{lamsal2022socially}).
2) We excluded papers where the intelligent system was not functionally implemented, but merely presented through conceptual or demonstrative prototypes to solicit user feedback (e.g., \cite{ashktorab2023SMEintheloop}).
3) In cases where multiple publications described the same system and evaluation, only the most representative version was retained.

Following this phase, 34 papers remained. To further expand the corpus, we conducted both backward and forward citation analysis on the current set, applying the same eligibility criteria to newly identified papers. This yielded a final corpus of 77 papers for subsequent analysis.

\subsubsection{Inclusion}

\begin{figure*}[htbp]
    \centering
    \includegraphics[width=0.8\textwidth]{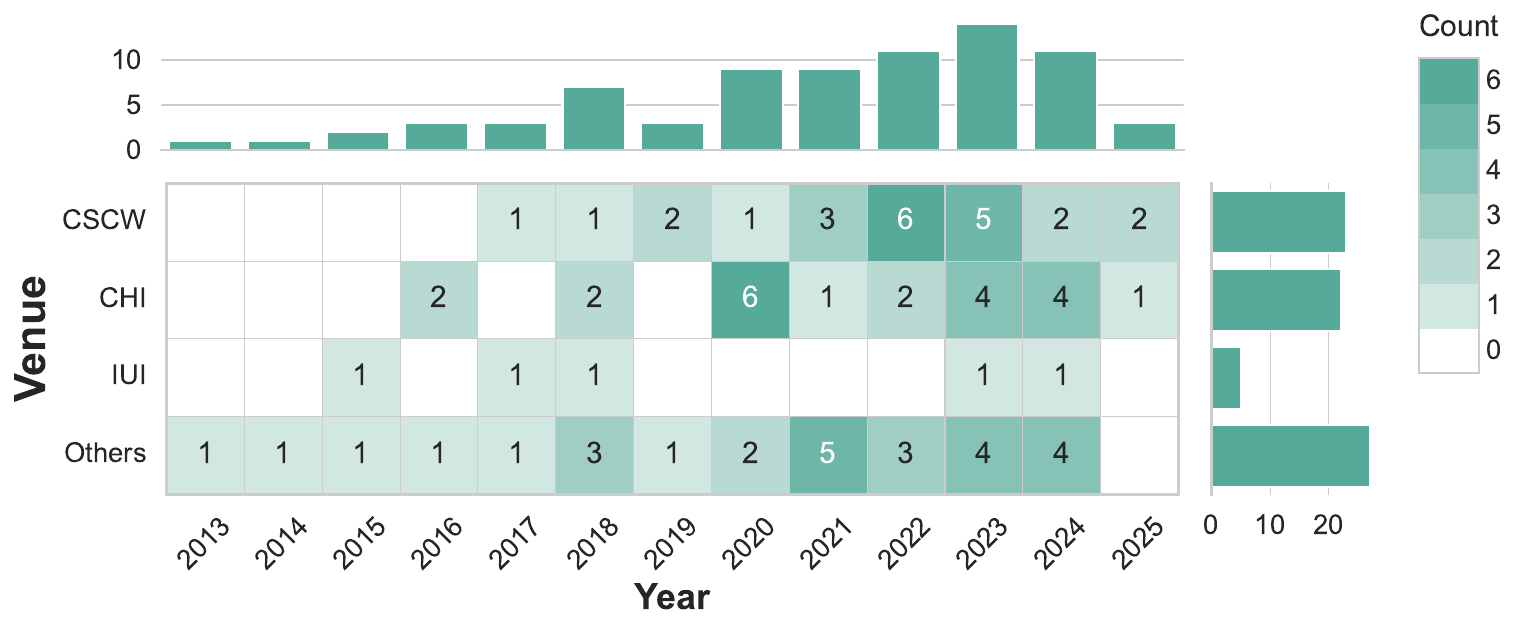}
    \caption{An overview of our collected papers in venues and years. Companion publications and extended abstracts are grouped with their corresponding main conference. Venues with no more  than three papers are  grouped under the category ``Others.''}
    \Description{}
    \label{fig: paper distribution overview}
\end{figure*}

The final corpus consists of 77 papers published between 2013 and 2025. The majority were published in top-tier HCI venues, with 29.9\% from CSCW and 28.6\% from CHI.  
As shown in \Cref{fig: paper distribution overview}, the number of publications has notably increased since 2020, reflecting a growing research interest in AI-infused systems for online communities.

\subsection{Data Analysis}

\label{sec:3.2}

Over the collected corpus of literature, we conduct a collaborative analysis to identify the key dimensions of our proposed RQs.
Two authors independently reviewed and coded all papers, using either an inductive or theory-driven approach depending on the dimension.
Specifically, for RQ1, we first coded the core challenges that the AI-infused systems aimed to address.
We then analyzed systems' design functionalities, together with whether the backend approach in the functions, their interaction scope (public or private), and the types of human support to AI involved in the design. 
For RQ2, we coded the evaluation metrics and associated data collection methods used to assess the impact of existing AI-infused systems. 
We followed a widely adopted taxonomy of data collection methods \cite{sharp2003interaction}, including: Observation (e.g., log analysis, behavioral traces), Questionnaire, and Interview.

For dimensions without established taxonomies, the two analysts first independently extracted and summarized relevant excerpts from the papers. These excerpts were then inductively coded, organized into themes to form an initial taxonomy following Wolfswinkel’s grounded theory approach \cite{wolfswinkel2013UsingGroundedTheory}.
Through discussion, they reconciled differences, refined the structure, and finalized the codebooks for each RQ. Based on the finalized codebooks, the two analysts re-reviewed the corpus, assigned codes to each paper, and resolved disagreements through discussion. The detailed results for each RQ are presented in \Cref{sec: RQ2}, and \Cref{sec: RQ3}, respectively.








\section{RQ1: How have AI‑infused systems been designed to address challenges in online communities?}
\label{sec: RQ2}


To design effective AI-infused systems for online communities, it is crucial to first identify the core challenges these systems are intended to address \cite{shi2023hci, lai2022human}. 
Our synthesis classifies these challenges into four overarching categories: 1) \textit{Contribution Barriers}, 2) \textit{Consumption Obstacles}, 3) \textit{Relational Challenges}, and 4) \textit{Moderation Limitations} (detailed codes and definitions in Appendix \ref{tab:codebook1}).

We align system functionalities with these challenges under four themes: (1) \textbf{Contribution Support}, (2) \textbf{Consumption Support}, (3) \textbf{Interaction Mediation}, and (4) \textbf{Moderation Support}. 
Each functionality is further annotated by: (1) back‑end approach (rule‑based, ML‑based, GenAI‑based), (2) human–AI interaction scope (public or private), and (3) how humans support AI in delivering the function. These dimensions reveal how functions leverage AI capabilities, are socially embedded, and involve human governance. 
Table \ref{tab:codebook2} summarizes the functionality codebook.

\begin{figure*}[htbp]
    \centering
    \includegraphics[width=0.85\textwidth]{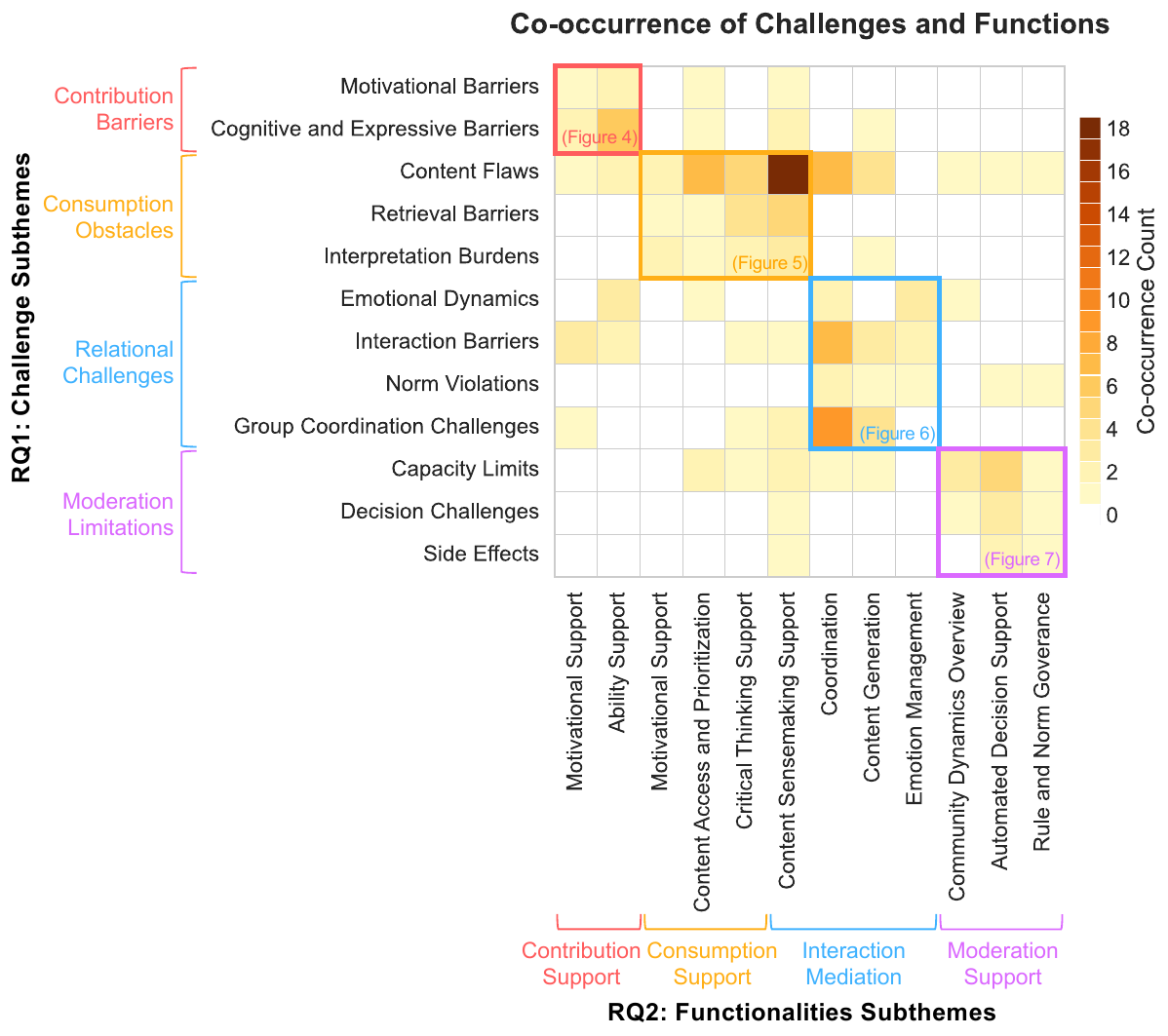}
    \caption{Co-occurrence heatmap of challenges (RQ1) and functionalities (RQ2) at the subtheme level.}
    \Description{}
    \label{fig: RQ1 RQ2 co-occurrence}
\end{figure*}

Beyond category construction, we map challenges and functionalities at the subtheme level and visualize their co‑occurrence in \Cref{fig: RQ1 RQ2 co-occurrence}.
This mapping surfaces broader patterns in the current design landscape, revealing dominant pairings, gaps, and trade‑offs, guiding designers in selecting appropriate approaches.
In the following sections, we first illustrate how functionalities (Table \ref{tab:codebook2}) address challenges (Appendix \ref{tab:codebook1}) within the same theme (i.e., boxed regions in Figure \ref{fig: RQ1 RQ2 co-occurrence}), and then explore 
cross-theme co-occurrences and tensions in Section \ref{sec: 7.1.2}, exploring how challenges in one domain may be addressed—or inadvertently exacerbated—by functionalities in another.

\begin{table}[htbp]
\centering
\smallscriptsize
\renewcommand{\arraystretch}{1.2}
\begin{tabular}{|p{0.15cm}| p{1.8cm}| p{6.87cm} |C{0.07cm}? C{0.07cm}? C{0.07cm}| C{0.07cm}? C{0.07cm}|C{0.07cm}? C{0.07cm}? C{0.07cm}? C{0.07cm}|}

\cline{4-12}
\multicolumn{1}{c}{} & 
\multicolumn{1}{c}{} & 
\multicolumn{1}{c}{} & 
\multicolumn{3}{|c|}{\multirow{1}{*}{\makecell{}{\parbox{8ex}{\textbf{Backend}}}}} & 
\multicolumn{2}{c|}{\textbf{Scope}} & 
\multicolumn{4}{c|}{\makecell{\textbf{Human}\\\textbf{Support}}} \\ 

\cline{4-12}





\multicolumn{1}{c}{}   & 
\multicolumn{1}{c}{} & 
\multicolumn{1}{c}{} & 
\multicolumn{1}{|C{0.07cm}?}{
\multirow{2}{*}{
        \makecell{\rotatebox{90}{\parbox{13.6ex}{Rule-based}}}
        }
}
&
\multirow{2}{*}{\makecell{\rotatebox{90}{\parbox{13.6ex}{ML-based}}}} & 
\multirow{2}{*}{\makecell{\rotatebox{90}{\parbox{13.6ex}{GenAI-based}}}} & 
\multirow{2}{*}{\makecell{\rotatebox{90}{\parbox{13.6ex}{Public}}}} & 
\multirow{2}{*}{\makecell{\rotatebox{90}{\parbox{13.6ex}{Private}}}} & 
\multirow{2}{*}{\makecell{\rotatebox{90}{\parbox{13.6ex}{Feeding}}}}  & 
\multirow{2}{*}{\makecell{\rotatebox{90}{\parbox{13.6ex}{Calibrating}}}}  & 
\multirow{2}{*}{\makecell{\rotatebox{90}{\parbox{13.6ex}{Customizing}}}} & 
\multirow{2}{*}{\rotatebox{90}{\parbox{13.6ex}{Configuring}}} 
\rule{0pt}{0.9cm}
\\

\cline{2-3}

\multicolumn{1}{c}{}   & \multicolumn{1}{|c|}{\textbf{Subtheme}}  & \multicolumn{1}{c|}{\textbf{Code}}  &  &  &  &  &  &  &  &  &  \\

\hline

\cellcolor{applepinknormal!20} & \multirow{5}{*}{\parbox{1.8cm}{Motivational\\Support}} & \cellcolor{applegrey!15} Encouragement Message \cite{do2022should, ito2022agent, liu2023coargue, nordberg2019designing, dyke2013enhancing, kumar2014triggering, wang2022understanding} & \Rule & \ML & & \PublicFunction & \PrivateFunction &  &  &  &  \\ \arrayrulecolor{applegrey!15}\cline{4-12}
\cellcolor{applepinknormal!20} &                      & Peer-Mediated Encouragement \cite{do2022should} & \Rule & & &  & \PrivateFunction &  &  &  &  \\ \arrayrulecolor{applegrey!15}\cline{4-12}
\cellcolor{applepinknormal!20} &                      & \cellcolor{applegrey!15} Contribution Impact Feedback \cite{liu2023coargue} & & \ML & &  & \PrivateFunction  &  &  &  &  \\ \arrayrulecolor{applegrey!15}\cline{4-12}
\cellcolor{applepinknormal!20} &                      & Growth Simulation \cite{seering2020takes} & \blank & \ML &  & \PublicFunction &  &  &  &  &  \\ \arrayrulecolor{applegrey!15}\cline{4-12}
\cellcolor{applepinknormal!20} &                      & \cellcolor{applegrey!15} First-Response Trigger \cite{wang2021cass} & \Rule & \ML & & \PublicFunction &  &  &  &  &  \\

\arrayrulecolor{applegrey!70}\cline{2-12}
\cellcolor{applepinknormal!20} & \multirow{7}{*}{\parbox{1.8cm}{Ability Support}}       & Self-Reflection Nudging \cite{yeo2024help} & \blank & \blank & \GenAIInvolved & & \PrivateFunction &  &  &  &  \\ \arrayrulecolor{applegrey!15}\cline{4-12}
\cellcolor{applepinknormal!20} &                       & \cellcolor{applegrey!15} Relevant Example Provision \cite{peng2020exploring, xia2022persua} & & \ML & &  & \PrivateFunction &  &  &  &  \\ \arrayrulecolor{applegrey!15}\cline{4-12}
\cellcolor{applepinknormal!20} &                       & Writing Recommendation and Guidance \cite{liu2023coargue, shin2022exploring, peng2020exploring, shin2023introbot, narain2020promoting, lee2020solutionchat, chen2022topsbot} & \Rule & \ML & &  & \PrivateFunction & \Training &  &  &  \\ \arrayrulecolor{applegrey!15}\cline{4-12}
\cellcolor{applepinknormal!20} &                       & \cellcolor{applegrey!15} Input Structure Analysis \cite{xia2022persua}  & & \ML & &  & \PrivateFunction &  &  &  &  \\ \arrayrulecolor{applegrey!15}\cline{4-12}
\cellcolor{applepinknormal!20} &                       & Writing Evaluation and Feedback \cite{peng2020exploring} & & \ML & &  & \PrivateFunction &  &  &  &  \\ \arrayrulecolor{applegrey!15}\cline{4-12}
\cellcolor{applepinknormal!20} &                       & \cellcolor{applegrey!15} Contextual Image Augmentation \cite{zhang2024mentalimager}  & \blank & \blank & \GenAIInvolved &  & \PrivateFunction &  &  &  & \Configuring \\ \arrayrulecolor{applegrey!15}\cline{4-12}
\multirow{-12}{*}{\rotatebox{90}{\textbf{Contribution Support}}\cellcolor{applepinknormal!20}} &                       & Descriptive Tagging \cite{shin2022exploring}  & & \ML & &  & \PrivateFunction &  &  &  &  \\
\arrayrulecolor{black}\hline


\cellcolor{applepurplenormal!20} & \multirow{2}{*}{\parbox{1.8cm}{Motivational\\Support}}  & \cellcolor{applegrey!15} Agent-Driven Dialogue \cite{peng2024designquizzer, zhang2024see} & \Rule & \ML & \GenAIInvolved &  & \PrivateFunction &  &  &  &  \\ \arrayrulecolor{applegrey!15}\cline{4-12}
\cellcolor{applepurplenormal!20} &                       & Positive Performance Feedback \cite{peng2024designquizzer}  & \Rule & & &  & \PrivateFunction &  &  &  &  \\
\arrayrulecolor{applegrey!40}\cline{2-12}
\cellcolor{applepurplenormal!20} & \multirow{3}{*}{\parbox{1.8cm}{Content Access and Prioritization}} & \cellcolor{applegrey!15} Content Ranking and Recommendation \cite{shin2022exploring, gao2023know, hoque2017cqavis, park2016supporting, tepper2018collabot}  & \Rule & \ML & & \PublicFunction & \PrivateFunction &  &    & \Customizing & \Configuring \\ \arrayrulecolor{applegrey!15}\cline{4-12}
\cellcolor{applepurplenormal!20} &                       &  Content Filtering \cite{ito2022agent, zhang2025coknowledge, hoque2017cqavis, park2016supporting} & \Rule & \ML & & \PublicFunction & \PrivateFunction &  &  &  & \Configuring \\ \arrayrulecolor{applegrey!15}\cline{4-12}
\cellcolor{applepurplenormal!20} &                       & \cellcolor{applegrey!15} Key Information Highlighting \cite{zhang2025coknowledge, liu2022planhelper, sun2016videoforest, kwon2015visohc, zhang2017wikum}  &  & \ML & & \PublicFunction & \PrivateFunction &  &  & \Customizing &  \\
\arrayrulecolor{applegrey!40}\cline{2-12}
\cellcolor{applepurplenormal!20} & \multirow{2}{*}{\parbox{1.8cm}{Critical Thinking\\Support}} &  Fact Verification and Explanation \cite{piccolo2021agents}  & \Rule & & & \PublicFunction & \PrivateFunction &  &  &  &  \\ \arrayrulecolor{applegrey!15}\cline{4-12}
\cellcolor{applepurplenormal!20} &                       & \cellcolor{applegrey!15} Facilitating Exposure to Diverse Perspectives \cite{govers2024ai, gao2018burst, tanprasert2024debate, zhang2024see, kim2021starrythoughts}  & \blank & \ML & \GenAIInvolved & \PublicFunction & \PrivateFunction & \Training & \Calibrating &  &  \\ 
\arrayrulecolor{applegrey!40}\cline{2-12}
\cellcolor{applepurplenormal!20} & \multirow{5}{*}{\parbox{1.8cm}{Content\\Sensemaking\\Support}}  &  Text Summarization \cite{kim2020bot, tepper2018collabot, wu2024comviewer,  peng2024designquizzer, zhang2018making, han2023redbot, lee2020solutionchat, zhang2017wikum}  & \Rule & \ML & \GenAIInvolved & \PublicFunction & \PrivateFunction & \Training &  & \Customizing & \\ \arrayrulecolor{applegrey!15}\cline{4-12}
\cellcolor{applepurplenormal!20} &                       & \cellcolor{applegrey!15} Solution Generation \cite{xu2017answerbot, wu2024comviewer, da2020crokage} & \blank & \ML & \GenAIInvolved &  & \PrivateFunction &  &  &  &  \\ \arrayrulecolor{applegrey!15}\cline{4-12}
\cellcolor{applepurplenormal!20} &                       &  Content Clustering \cite{liu2023coargue, zhang2025coknowledge, hoque2017cqavis}  & & \ML & &  & \PrivateFunction &  &  &  &  \\ \arrayrulecolor{applegrey!15}\cline{4-12}
\cellcolor{applepurplenormal!20} &                       & \cellcolor{applegrey!15}Structural Visualization \cite{ito2022agent, zhang2025coknowledge, perez2018collaborative, hoque2015convisit, liu2022planhelper, aiyer2023taming, sun2016videoforest, zhang2017wikum}  & \blank & \ML & \GenAIInvolved & \PublicFunction & \PrivateFunction &  &  & \Customizing &  \\ \arrayrulecolor{applegrey!15}\cline{4-12}
\multirow{-12}{*}{\rotatebox{90}{\textbf{Consumption Support}}\cellcolor{applepurplenormal!20}} &                       &  Analytical Visualization \cite{gao2018burst, liu2023coargue, zhang2025coknowledge, rifat2024combating, wu2024comviewer, peng2024designquizzer, liu2022planhelper, kim2021starrythoughts, park2016supporting, fu2018t, chen2022topsbot, sun2016videoforest, cao2023visdmk} & \Rule & \ML & \GenAIInvolved & \PublicFunction & \PrivateFunction &  & \Calibrating & \Customizing &  \\
\arrayrulecolor{black}\hline

\cellcolor{applegreenlight!20} & \multirow{8}{*}{\parbox{1.8cm}{Coordination}}         & \cellcolor{applegrey!15}Homophily Grouping \cite{wang2020jill, wang2022understanding} & \blank & \ML & \blank & \PublicFunction &  &  &  &  &  \\ \arrayrulecolor{applegrey!15}\cline{4-12}
\cellcolor{applegreenlight!20} &                      &  Conversation Activation \cite{shin2021blahblahbot, shin2023introbot, niksirat2023potential, wang2022understanding}  & \Rule & \ML & \blank & \PublicFunction & \PrivateFunction &  &  &  &  \\ \arrayrulecolor{applegrey!15}\cline{4-12}
\cellcolor{applegreenlight!20} &                      & \cellcolor{applegrey!15} Participation Balancing \cite{bagmar2022analyzing, kim2020bot, do2022should, do2023inform, shin2023introbot, kim2021moderator, do2023err} & \Rule & \blank & \blank & \PublicFunction & \PrivateFunction &  & \Calibrating &  & \Configuring \\ \arrayrulecolor{applegrey!15}\cline{4-12}
\cellcolor{applegreenlight!20} &                      &  Task Monitoring \cite{toxtli2018understanding}  & \blank & \ML & \blank &  & \PrivateFunction &   &   &   & \Configuring    \\ \arrayrulecolor{applegrey!15}\cline{4-12}
\cellcolor{applegreenlight!20} &                      & \cellcolor{applegrey!15} Perspective-Taking Prompting \cite{shin2022chatbots} & \Rule & \blank & \blank & \PublicFunction &  &   &  &  &  \\ \arrayrulecolor{applegrey!15}\cline{4-12}
\cellcolor{applegreenlight!20} &                      &  Allyship Facilitation \cite{ueda2021cyberbullying}  & & & & \PublicFunction &  &   &  &  & \\ \arrayrulecolor{applegrey!15}\cline{4-12}
\cellcolor{applegreenlight!20} &                      & \cellcolor{applegrey!15} Topic Steering \cite{govers2024ai, bagmar2022analyzing, ito2022agent, shin2021blahblahbot, nordberg2019designing, gimpel2024digital, dyke2013enhancing, shin2023introbot, kim2021moderator, niksirat2023potential, han2023redbot, lee2020solutionchat} & \Rule & \ML & \GenAIInvolved & \PublicFunction & \PrivateFunction & \Training  &  &  & \\ \arrayrulecolor{applegrey!15}\cline{4-12}
\cellcolor{applegreenlight!20} &                      &  Time Management \cite{lee2013analyzing, kim2020bot, benke2020chatbot, shin2023introbot, kim2021moderator} & \Rule & \blank & \blank & \PublicFunction &  &   &  &  &  \\
\arrayrulecolor{applegrey!40}\cline{2-12}
\cellcolor{applegreenlight!20} & \multirow{6}{*}{\parbox{1.8cm}{Content\\Generation}}   & \cellcolor{applegrey!15} Providing Suggestions \cite{shin2022chatbots, niksirat2023potential, han2023redbot, avula2022effects, gong2024cosearchagent} & \Rule & \blank & \GenAIInvolved & \PublicFunction & \PrivateFunction &  &  &  &  \\ \arrayrulecolor{applegrey!15}\cline{4-12}
\cellcolor{applegreenlight!20} &                      & AI-Generated Social Support \cite{murgia2016among, wang2021cass, kumar2014triggering, piccolo2021agents}  & \Rule & \blank & \GenAIInvolved & \PublicFunction &  &   &  &  &     \\ \arrayrulecolor{applegrey!15}\cline{4-12}
\cellcolor{applegreenlight!20} &                      & \cellcolor{applegrey!15} Message Restating \cite{dyke2013enhancing}  & & \ML & & \PublicFunction &  &   &  &  &  \\ \arrayrulecolor{applegrey!15}\cline{4-12}
\cellcolor{applegreenlight!20} &                      & Argument Support/Refutation \cite{hadfi2021argumentative, chiang2024enhancing, kumar2014triggering} & \Rule & \ML & \GenAIInvolved & \PublicFunction &  &   &  &  &  \\ \arrayrulecolor{applegrey!15}\cline{4-12}
\cellcolor{applegreenlight!20} &                      & \cellcolor{applegrey!15} Information Retrieval \cite{de2024assessing, gong2024cosearchagent, avula2019embedding, avula2018searchbots, hecht2012searchbuddies, avula2022effects} & \Rule & \ML & \GenAIInvolved & \PublicFunction &  &   &  &  &  \\ \arrayrulecolor{applegrey!15}\cline{4-12}
\cellcolor{applegreenlight!20} &                      & Evidence Provision \cite{gong2024cosearchagent, hecht2012searchbuddies, govers2024ai}  & \blank & \ML & \GenAIInvolved & \PublicFunction &  &   &  &  &  \\
\arrayrulecolor{applegrey!40}\cline{2-12}
\cellcolor{applegreenlight!20} & \multirow{3}{*}{\parbox{1.8cm}{Emotion\\Management}}   & \cellcolor{applegrey!15} Emotion State Delivery \cite{benke2020chatbot, peng2019gremobot}  & & \ML & & \PublicFunction &  &   &  &  &  \\ \arrayrulecolor{applegrey!15}\cline{4-12}
\cellcolor{applegreenlight!20} &                      & Positive Social Cueing \cite{benke2020chatbot, peng2019gremobot, narain2020promoting}  & \Rule & & & \PublicFunction & \PrivateFunction &  &  &  &  \\ \arrayrulecolor{applegrey!15}\cline{4-12}
\multirow{-17}{*}{\rotatebox{90}{\textbf{Interaction Mediation}}\cellcolor{applegreenlight!20}} &                      & \cellcolor{applegrey!15} Negative Content Pruning \cite{benke2020chatbot}  &\Rule & & & \PublicFunction &  &   &  &  &   \\
\arrayrulecolor{black}\hline

\cellcolor{appleyellowlight!20} & \multirow{2}{*}{\parbox{1.8cm}{Community Dynamics Overview}} &  Visualizing Community Activity \cite{hee2024brinjal,choi2023convex, park2016supporting, fu2018t, kwon2015visohc}  & \Rule & \ML & &  & \PrivateFunction &  &  &  &  \\ \arrayrulecolor{applegrey!15}\cline{4-12}
\cellcolor{appleyellowlight!20} &                      & \cellcolor{applegrey!15} Summarizing Demographics \cite{wang2020jill} &  & \ML &  & \PublicFunction &  &   &  &  &  \\
\arrayrulecolor{applegrey!40}\cline{2-12}
\cellcolor{appleyellowlight!20} & \multirow{2}{*}{\parbox{1.8cm}{Automated\\Decision Support}} &  Antisocial Behavior Forecasting \cite{choi2023convex, schluger2022proactive}  & & \ML &  &  & \PrivateFunction &  &  &  &  \\ \arrayrulecolor{applegrey!15}\cline{4-12}
\cellcolor{appleyellowlight!20} &                      & \cellcolor{applegrey!15} Content Flagging \cite{hee2024brinjal, rifat2024combating, chandrasekharan2019crossmod, he2023cura, jahanbakhsh2023exploring, lai2022human}  &  & \ML & & \PublicFunction & \PrivateFunction & \Training & \Calibrating &   & \Configuring \\
\arrayrulecolor{applegrey!40}\cline{2-12}
\cellcolor{appleyellowlight!20} & \multirow{3}{*}{\parbox{1.8cm}{Rule and Norm\\Governance}} &  Moderation Rule Analysis \cite{song2023modsandbox}  & & \ML &  &  & \PrivateFunction &   &   &   & \Configuring \\ \arrayrulecolor{applegrey!15}\cline{4-12}
\cellcolor{appleyellowlight!20} &                      & \cellcolor{applegrey!15} Norm Violation Alerting \cite{bagmar2022analyzing}  & \Rule & &  & \PublicFunction &  &   &  &  &  \\ \arrayrulecolor{applegrey!15}\cline{4-12}
\multirow{-7}{*}{\rotatebox{90}{\textbf{Moderation Support}}\cellcolor{appleyellowlight!20}} &                      &  Appeal Reflection Prompting \cite{atreja2024appealmod}  & \Rule & & &  & \PrivateFunction &  &  &  &  \\
\arrayrulecolor{black}

\hline

\end{tabular}%
\caption{Functionalities of AI-infused systems identified in our corpus, categorized into four types of support: \textbf{Contribution, Consumption, Mediation}, and \textbf{Moderation}. Each function is further annotated with: (1) backend techniques, (2) public or private interaction scope, and (3) types of human support.}
\label{tab:codebook2}
\end{table}



\normalsize
\subsection{Contribution Support}

Online communities depend on users to generate fresh content and engage in timely interactions \cite{malinen2015understanding, armstrong2009real}. A rich inventory of high-quality UGC is vital for keeping the community vibrant and valuable to its members \cite{chen2011moderated,  engel2015collective}. As such, supporting user contribution has long been a central goal of online community platforms \cite{malinen2015understanding, ellis2004community, lampe2010motivations}, yet users often face two main challenges: \textit{motivational barriers} and \textit{cognitive or expressive barriers}.

\subsubsection{Motivational Support}
\label{sec: 5.1.1}
This subsection examines functionalities designed to address \textit{motivational barriers}.
These systems leverage AI’s ability to rapidly process large volumes of messages to infer conversational context and act as a perceived social actor \cite{nass1994computers}.
\yh{Consequently, they deliver timely, consistent interventions that serve as social triggers to elicit accountability, operating at a scale unattainable by human facilitators \cite{do2022should, ito2022agent}.}

To address \textit{low confidence in contributing},
a common approach is the delivery of \textbf{encouragement messages} \cite{do2022should, ito2022agent, liu2023coargue, nordberg2019designing, dyke2013enhancing, kumar2014triggering, wang2022understanding}.
\yh{By identifying under-contributing members, systems target them with praise or invitations to participate.}
Beyond direct prompts, some systems also attempt to \textbf{mobilize peer encouragement} \yh{by privately requesting active members to engage quiet users \cite{do2022should}.}
While such encouragement can foster engagement, it also entails risks: public prompts to individuals may cause embarrassment, feelings of surveillance, and subsequent withdrawal \cite{do2022should}. 
Moreover, simple encouragement messages may lead users to take the easiest path -- posting more frequently without improving the quality of their ideas -- thereby degrading content \cite{do2023inform}.

\yh{Another strategy focuses on validating user effort through \textbf{contribution feedback} \cite{liu2023coargue}. By quantifying how a post expands the community’s knowledge repertoire, tools like CoArgue \cite{liu2023coargue} visualize the tangible value of user inputs.} Dyke et al. \cite{dyke2013enhancing} found that public AI feedback can increase contributions but risks shifting interaction from human–human to human–AI.

Other systems help \textit{maintain contribution engagement} through \textbf{playful nurturing interactions}—e.g., Seering et al. \cite{seering2020takes} deployed BabyBot, where members “raised” it from “birth” to “adulthood” by contributing ideas. 
\yh{This setting gamifies interaction and sustains interest}; even AI's nonsensical messages did not reduce engagement but often sparked playful exchanges.
Additionally, some systems (e.g., CASS \cite{wang2021cass}) lower \textit{anticipated contribution burden} by \textbf{auto-generating first responses}. 
\yh{By detecting unanswered posts and initiating the first reply, the AI reduces the social pressure on users to break the silence, thereby facilitating follow-up contributions.}

\subsubsection{Ability Support}

Even when users are emotionally engaged, \textit{cognitive and expressive barriers} can hinder their ability to produce high‑quality contributions.
While traditional human training can improve performance, it is time‑intensive and prone to high attrition \cite{geraedts2014long, lederman2014moderated}. 
\yh{AI‑infused systems offer a scalable alternative by providing adaptive scaffolding support throughout the content creation process.}

\yh{To overcome difficulties in ideation (such as \textit{locating contribution angles} and \textit{organizing thoughts}), systems leverage AI's retrieval and generative capabilities to provide semantically relevant content.}
Common strategies include providing \textbf{writing recommendations} \cite{liu2023coargue, shin2022exploring, peng2020exploring, shin2023introbot, narain2020promoting, lee2020solutionchat, chen2022topsbot}, \yh{which range from suggesting high-level topics to generating specific message candidates,} or presenting \textbf{topic-aligned well-crafted examples} \cite{xia2022persua, peng2020exploring} as benchmarks for imitation.
Beyond direct suggestions, systems like HelpMeReflect \cite{yeo2024help} utilize LLMs' language proficiency to deliver \textbf{self-reflection nudges}, prompting users to consider diverse perspectives while formulating their comments in online deliberation.
However, extensive reliance on AI-generated content risks saturating communities with homogeneous inputs, potentially eroding the sincerity and originality of human interaction.

Once a draft is formulated, \yh{systems act as automated critics to facilitate refinement. Tools like Persua} \cite{xia2022persua} provide \textbf{input structure analysis}, which visually decompose argumentative structures and rhetorical strategies of a message for users with \textit{limited argumentation skills}.
Others offer \textbf{writing evaluation and feedback} to improve \textit{language tailoring} or \textit{social support quality}; for example, MepsBot \cite{peng2020exploring} assesses the emotional and informational quality of input text and offers tailored suggestions for improvement. 
Although these features effectively leverage ML models to automate linguistic and structural analysis, algorithmic inaccuracies remain. Furthermore, users may feel frustrated by the sense of being “judged” by an AI, which can potentially deter future contributions \cite{o2017design}.

After the main text is completed, systems further enhance the visibility and clarity through \textbf{descriptive tag generation} \cite{shin2022exploring} 
(e.g.,  extracting contextual keywords) 
and \textbf{contextual image augmentation}, transforming textual inputs into more accessible, multimodal content. For example, MentalImager \cite{zhang2024mentalimager} employs text-to-image models to create emotion-relevant images that assist self-disclosure, \yh{though ensuring the semantic relevance of such generated images remains a challenge.}

\subsubsection{Design Space}
\begin{figure*}[htbp]
    \centering
    \includegraphics[width=0.75\textwidth]{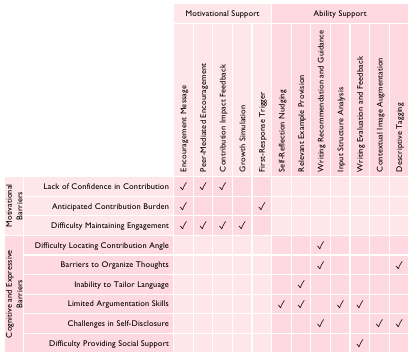}
    \caption{\yh{Design space for AI-infused system to support content contribution.} 
    }
    \Description{Design space for content contribution.}
    \label{fig: design space contribution}
\end{figure*}
\Cref{fig: design space contribution} illustrates the design landscape for contribution support. 
First, distinct strategies can address the same challenge, each with unique trade-offs.
\yh{For instance, to mitigate lack of confidence, designers can employ either system-generated or peer-mediated encouragement.
Peer mediation fosters inclusiveness and natural interaction \cite{do2022should}, potentially stimulating community-driven self-mediation; however, it risks causing social friction if users feel annoyed by peers \cite{do2022should}.}
In contrast, system-generated  messages, especially when private, reduce the feeling of embarrassment and are more suitable for individual settings. 
\yh{Second, certain functionalities demonstrate versatility by addressing multiple barriers simultaneously.} For example, writing recommendation and guidance can alleviate both ideation and refinement difficulties depending on their specific implementation~\cite{liu2023coargue, peng2020exploring, shin2023introbot}. 
\yh{Third, the design space highlights a clear alignment between support types and barrier categories, which} implies that designers can combine those functions to create composite pipelines that support users through different stages of contribution.

\subsection{Consumption Support}

Compared to contribution, content consumption is a more pervasive activity in online communities, as nearly all members engage in it. Effective absorption and interpretation of existing content are essential not only for realizing the community’s value \cite{armstrong2009real, lampe2010motivations}, but also for enabling further participation \cite{preece2009reader, kittur2008harnessing}.
Yet users often encounter challenges during this process.
First, the nature of user‑generated content (UGC)—shaped by diverse contributors and the absence of gatekeeping—can result in various \textit{content flaws}. Second, they also encounter difficulties while retrieving relevant information (\textit{retrieval barriers}). Finally, even when users successfully retrieve high-quality content, they may still face challenges in interpreting the information (\textit{interpretation burdens}). The following sections outline functionalities designed to address these three categories of challenges.

\subsubsection{Motivational Support}
\yh{Given that consumption is often passive, users frequently \textit{struggle to maintain cognitive engagement}, especially when facing dense material.}
Systems address this by leveraging AI’s scalability and its ability to act as a human-like collaborator (similar to section \ref{sec: 5.1.1}).
One approach involves \textbf{agent-driven dialogue} \cite{zhang2024see, peng2024designquizzer}, \yh{where agents (e.g., DesignQuizzer \cite{peng2024designquizzer}) pose questions to prompt reflection, thereby shifting users from passive absorption to interactive co-construction. }
Another strategy is \textbf{providing positive feedback} during reading to help users maintain cognitive engagement and motivation throughout the process \cite{peng2024designquizzer}.

\subsubsection{Content Access and Prioritization}

The \textit{sheer volume} of UGC—often \textit{uneven in quality} and \textit{off‑topic}—makes finding valuable information challenging. 
\yh{AI-infused systems address this by leveraging NLP techniques (e.g., semantic analysis, summarization) to automate the curation pipeline, effectively shifting the discovery burden from human users to algorithms.

At a macro level, systems employ \textbf{content ranking and recommendation} \cite{shin2022exploring, gao2023know, hoque2017cqavis, park2016supporting, tepper2018collabot} to prioritize high-value material. A key advantage here is independence from user labor: unlike crowdsourced sorting (e.g., upvotes) which requires active community participation, AI techniques like topic modeling (e.g., Collabot \cite{tepper2018collabot}) operate autonomously, ensuring content visibility even in smaller user bases \cite{park2016supporting}.}
\yh{At a micro level, systems facilitate granular navigation through} \textbf{content filtering} \cite{ito2022agent, zhang2025coknowledge, hoque2017cqavis, park2016supporting}, which allows users to selectively view content based on various dimensions (e.g., topic, source). 
Additionally, \yh{systems like Wikum \cite{zhang2017wikum} offer \textbf{key information highlighting} that automatically detect salient insights, enabling users to rapidly skim and comprehend complex discussions \cite{zhang2025coknowledge, liu2022planhelper, sun2016videoforest, kwon2015visohc, zhang2017wikum}}.

\subsubsection{Critical Thinking Support}

To mitigate \textit{misinformation}, some systems offer \textbf{fact-checking with explanations} for potentially misleading content. 
For example, Piccolo et al. \cite{piccolo2021agents} \yh{leveraged AI agents to not only flag errors but also privately explain the rationale.}
Another common challenge is the \textit{filter bubble effect} - intellectual isolation driven by algorithmic personalization.
A common countermeasure is \textbf{facilitating exposure to diverse perspectives} \yh{through simulated dialogue \cite{govers2024ai, gao2018burst, tanprasert2024debate, zhang2024see, kim2021starrythoughts}. 
By leveraging LLMs' capacity to simulate distinct personas, tools like debate chatbots \cite{tanprasert2024debate} or multi-perspective agents \cite{zhang2024see} intentionally adopt different stances to prompt reflection.
While human-to-human debate offers similar benefits, finding trustworthy partners with opposing views is difficult to scale \cite{tanprasert2024debate}.}
AI addresses this limitation by serving as a safe, nonjudgmental interlocutor, enabling such interactions at scale without the fear of offending others.
However, LLMs are not without drawbacks—they may generate inaccurate character representations or biased outputs stemming from their training data \cite{gao2018burst, tanprasert2024debate}.

\subsubsection{Content Sensemaking Support}

\yh{To support making sense of complex information, AI systems transform raw content into digestible formats through textual synthesis and visual presentation.

For textual synthesis,}
systems can \textbf{generate thread or post-level summaries} \cite{kim2020bot, tepper2018collabot, wu2024comviewer,  peng2024designquizzer, zhang2018making, han2023redbot, lee2020solutionchat, zhang2017wikum} to help users manage the \textit{fast-paced information flow},
or \textbf{produce direct solutions to user queries} \cite{xu2017answerbot, wu2024comviewer, da2020crokage} by synthesizing relevant content, which may be \textit{incomplete} when presented in isolation.
Others perform \textbf{content clustering} \cite{liu2023coargue, zhang2025coknowledge, hoque2017cqavis}\yh{—such as CoKnowledge's \cite{zhang2025coknowledge} merging of semantically similar comments—}to reduce \textit{redundancy} and streamline \textit{navigation of related content}. 
However, users may distrust or feel overly guided by such outputs and \yh{feel compelled to verify} original discussions \cite{liu2022planhelper, zhang2018making}. Moreover, when summaries are presented by agents like GroupfeedBot \cite{kim2020bot} as long messages to the group, they can disrupt conversational flow and occupy excessive screen space.

\yh{For visual presentation, systems reveal latent patterns through two main approaches.}
\textbf{Structural visualizations} \cite{ito2022agent, zhang2025coknowledge, perez2018collaborative, hoque2015convisit, liu2022planhelper, aiyer2023taming, sun2016videoforest, zhang2017wikum} reveal the organization within \textit{poorly structured} content; \yh{for instance, SOCIO \cite{perez2018collaborative} leverages relationship extraction to generate concept diagrams.}
\textbf{Analytical visualizations}  \cite{gao2018burst, liu2023coargue, zhang2025coknowledge, rifat2024combating, wu2024comviewer, peng2024designquizzer, liu2022planhelper, kim2021starrythoughts, park2016supporting, fu2018t, chen2022topsbot, sun2016videoforest, cao2023visdmk}  interpret \textit{thematically complex content} through specific interpretive lenses (such as stance or emotion) \yh{by visually rendering these AI-detected attributes}.

\subsubsection{Design Space}
\begin{figure*}[htbp]
    \centering
    \includegraphics[width=0.8\textwidth]{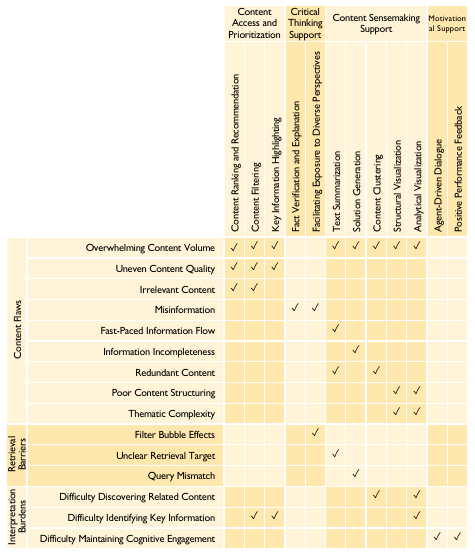}
    \caption{\yh{Design space for AI-infused system to support content consumption.} 
    }
    \Description{Design space for content consumption.}
    \label{fig: design space consumption}
\end{figure*}
\Cref{fig: design space consumption} outlines the design space for content consumption. 
\yh{First,} critical thinking supports specifically target \textit{misinformation} and \textit{filter bubble effects}. Unlike other barriers, these challenges are inherently ``deceptive,'' \yh{thus requiring targeted interventions that prompt active reflection rather than passive organization}.
\yh{Second, designers face a fundamental trade-off between presenting original content (via structuring or highlighting)} and presenting summarized or filtered content.
Presenting original content---albeit with structural support---offers users a greater opportunity for self-verification and \yh{detecting inherent flaws}. In contrast, summarizing or filtering contents, while more concise and digestible, may introduce misleading generated content or possible selection bias caused by the algorithm, which calls for careful considerations.

\subsection{Interaction Mediation}

Online communities are inherently social spaces, where participation relies on both individual skills and interpersonal dynamics  \cite{preece2009reader, kraut2012building}. In multi-user settings, healthy human-human interactions are critical for social cohesion and collective outcomes \cite{olson2000distance, haythornthwaite2005social}. However, the complexity of online relationships and the nature of online interaction often give rise to relational challenges: \textit{poor emotional dynamics}, \textit{interaction barriers}, \textit{norm violations}, and \textit{group coordination challenges}.

\subsubsection{Coordination}
This category includes functions in which AI-infused systems act as coordinators in group collaborations without contributing substantive content. 
\yh{To establish a cohesive foundation,
AI systems leverage user profiling capabilities to} \textbf{group members based on shared characteristics} \cite{wang2020jill, wang2022understanding}. For example, SAMI \cite{wang2022understanding} utilized Named Entity Recognition to extract attributes like locations and hobbies for grouping, \yh{though potentially limiting user agency in building social connections.}
Once groups are formed,\yh{systems act as social actors to} \textbf{initiate interactions} \cite{shin2021blahblahbot, shin2023introbot, niksirat2023potential, wang2022understanding} by suggesting topics or sending ice-breaker posts to spark engagement.

During ongoing discussions, AI can monitor participation dynamics. \yh{A common function is \textbf{promoting more balanced involvement} by identifying under-contributing or dominant members \cite{bagmar2022analyzing, kim2020bot, do2022should, do2023inform, shin2023introbot, kim2021moderator, do2023err}.
However, relying on simplistic measures like message counts (e.g., DebateBot \cite{kim2021moderator}) can incentivize non-substantive contributions.
Beyond frequency, systems address \textit{low social awareness} by \textbf{prompting perspective-taking} \cite{shin2022chatbots}, where the AI collects and presents diverse viewpoints to prompt reflection, lowering the social pressure of human moderators.
In extreme cases like} \textit{cyberbullying}, AI can even involve bystanders and prompt them to align with the victim, \textbf{fostering protective alliances} \cite{ueda2021cyberbullying}. Yet such interventions may infringe on bystander autonomy and discourage sustained system use.

\yh{Finally, to ensure group efficiency in certain tasks,} AI systems can \textbf{steer the direction} \cite{govers2024ai, bagmar2022analyzing, ito2022agent, shin2021blahblahbot, nordberg2019designing, gimpel2024digital, dyke2013enhancing, shin2023introbot, kim2021moderator, niksirat2023potential, han2023redbot, lee2020solutionchat} of discussions to keep collaboration focused and aligned with collective goals, such as \textit{reaching consensus}.
Some systems also support \textbf{time management} \cite{lee2013analyzing, kim2020bot, benke2020chatbot, shin2023introbot, kim2021moderator} by issuing reminders, managing discussion pacing, or prompting breaks when needed, \yh{although overly rigid controls may undermine users’ sense of agency.
Additionally, systems reduce coordination overhead through \textbf{task monitoring} \cite{toxtli2018understanding}, employing intent detection from user utterances to automatically track progress and allocate responsibilities.}

\subsubsection{Content Generation}
This category covers functions in which AI‑infused systems actively contribute by generating content or input that directly shapes community interactions. One key function is \textbf{providing suggestions} \cite{shin2022chatbots, niksirat2023potential, han2023redbot, avula2022effects, gong2024cosearchagent} based on group context to help achieve shared goals or enhance collaborative efficiency, \yh{such as proposing middle-ground solutions to resolve disagreements, though this risks evoking a sense of surveillance.}
AI agents may also \textbf{generate informational or emotional support messages} \cite{murgia2016among, wang2021cass, kumar2014triggering, piccolo2021agents} to address \textit{delays in peer‑provided support}.
For example, CASS \cite{wang2021cass} employed a generation model to provide emotional support to overlooked posts.

In addition to producing original content, some systems operate by building on others’ messages—such as \textbf{restating a member’s comment} \cite{dyke2013enhancing} \yh{to reinforce valid points}
or \textbf{explicitly supporting or contesting others' viewpoints} \cite{hadfi2021argumentative, chiang2024enhancing, kumar2014triggering}. 
\yh{For instance, agents have been deployed to play a devil's advocate role by arguing against previous discussions, thereby stimulating critical thinking and reducing over-reliance on AI \cite{chiang2024enhancing}.}

Furthermore, AI \yh{augments discussions by} \textbf{retrieving contextually relevant information} \cite{de2024assessing, gong2024cosearchagent, avula2019embedding, avula2018searchbots, hecht2012searchbuddies, avula2022effects} to advance collaborative tasks, \yh{often by reactively or proactively surfacing search results related to user queries.
However, proactive interventions must be calibrated to avoid undermining user agency.}
To further increase the credibility of generated content, systems support \textbf{attaching relevant evidence or sources} \cite{gong2024cosearchagent, hecht2012searchbuddies, govers2024ai}, \yh{where the AI (e.g., CoSearchAgent \cite{gong2024cosearchagent}) verifies its outputs by explicitly citing references or summarizing retrieved data to support its claims. }

\subsubsection{Emotion Management}
AI-infused systems can also regulate the emotional climate of online communities through a variety of mechanisms. Some perform sentiment analysis or tone detection based on the group messages to \textbf{assess and surface the group’s collective emotional state} \cite{benke2020chatbot, peng2019gremobot} to enhance members’ emotional awareness. 
Others foster a more positive atmosphere by \textbf{posting uplifting social cues} \cite{benke2020chatbot, peng2019gremobot, narain2020promoting} such as encouraging language or soothing imagery.
When emotional states deteriorate, certain systems intervene by \textbf{pruning negative or harmful content}, \eg, deleting the chat forcefully or upon the team's approval \cite{benke2020chatbot}.

\subsubsection{Design Space}
\begin{figure*}[htbp]
    \centering
    \includegraphics[width=0.95\textwidth]{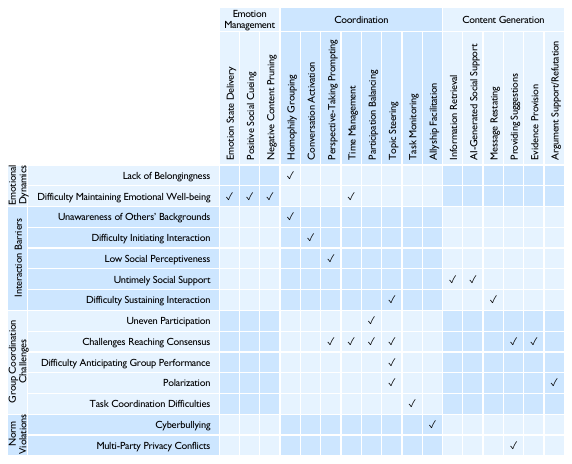}
    \caption{\yh{Design space for AI-infused system to support interaction mediation. }
    }
    \Description{Design space for interaction mediation.}
    \label{fig: design space mediation}
\end{figure*}
\Cref{fig: design space mediation} outlines the design space for group interaction mediation, which also reveals several design rationales.
First, addressing coordination challenges, especially reaching consensus, \yh{relies on a multi-layered intervention strategy. At the social level, systems facilitate the exchange of perspectives through mechanisms like perspective-taking and participation balancing. This is complemented by structural control over the discussion agenda (e.g., topic steering) and direct content contributions, where the AI takes initiative to provide suggestions. These strategies are not mutually exclusive but can be combined dynamically depending on group dynamics and progress.}
Secondly, there might be a causal relationship between different challenges, \yh{so functionalities addressing upstream issues can mitigate downstream friction}. For example, a lack of belongingness may cause uneven participation, and thus the functions like homophily grouping may eventually help to reach a more balanced contribution in group coordination tasks.

\subsection{Moderation Support}

Moderation is a core mechanism in online communities, typically involving the detection, alerting, and removal of inappropriate or harmful content and behavior \cite{jhaver2019does}. Effective moderation is essential for upholding community norms, protecting users from harm, and maintaining the overall quality of UGC \cite{gillespie2018custodians, lampe2004slash}. However, as communities scale in size and complexity, moderation systems face limitations, including \textit{capacity limits}, \textit{decision challenges}, and \textit{side effects} that affect both moderators and community members.

\subsubsection{Community Dynamics Overview}
A fundamental constraint of content governance is the \textit{excessive moderation load}, particularly in large-scale communities.
To alleviate it, AI‑infused systems \yh{can extract salient information based on key metrics, enabling moderators to quickly grasp community dynamics, thus accelerating the moderation process.}
One commonly observed function is the \textbf{visualization of community activity} \cite{hee2024brinjal,choi2023convex, park2016supporting, fu2018t, kwon2015visohc}.
For example, ConVex \cite{choi2023convex} calculates toxicity scores of the messages, and provides an analytical dashboard for moderators to locate problematic content efficiently.
\yh{Complementing this behavioral view, systems also support \textbf{demographic summarization} \cite{wang2020jill} by extracting and aggregating} demographic or identity-related information to help moderators contextualize user participation.

\subsubsection{Automated Decision Support}
Human moderators are often equipped with AI systems that provide automated decision support. 
\yh{In a proactive capacity, systems \textbf{perform antisocial behavior forecasting} \cite{choi2023convex, schluger2022proactive} by analyzing activity histories to predict the likelihood of conflict escalation before incidents occur, though such models often lack causal reasoning and may propagate historical biases.
Complementing this, \textbf{content flagging} \cite{hee2024brinjal, rifat2024combating, chandrasekharan2019crossmod, he2023cura, jahanbakhsh2023exploring, lai2022human} serves a reactive function by surfacing potential violations-either through indicative metrics or direct text classification-for moderator review. 
While systems like Cura \cite{he2023cura} demonstrate how such automation increases coverage and supports personalization,} they also raises ethical concerns regarding echo chambers and a lack of contextual sensitivity required for nuanced judgment.
Nonetheless, both approaches help narrow the scope of material requiring moderator attention, thereby reducing workload and enabling more targeted interventions.

\subsubsection{Rule and Norm Governance}
\yh{To mitigate errors in initial rule definitions of automated moderation tools}, AI systems support \textbf{rule analysis and enforcement}. 
For instance, ModSandbox \cite{song2023modsandbox} employs semantic similarity to \yh{visualize and prevent misclassifications in pre-configured rules.
Shifting to active interactions, AI facilitates \textbf{norm violation alerting} \cite{bagmar2022analyzing}. By detecting non-inclusive language via keyword matching, agents can privately alert users and suggest alternative phrasing.
Finally, to manage the influx of insincere appeals regarding moderation decisions,} systems provide \textbf{appeal reflection prompting} \cite{atreja2024appealmod}.
\yh{By introducing structural friction into the process, tools like AppealMod \cite{atreja2024appealmod} guide users to reflect on their behavior before submitting, thereby filtering out low-effort appeals and protecting moderators from harassment.}

\subsubsection{Design Space}
\begin{figure*}[htbp]
    \centering
    \includegraphics[width=0.6\textwidth]{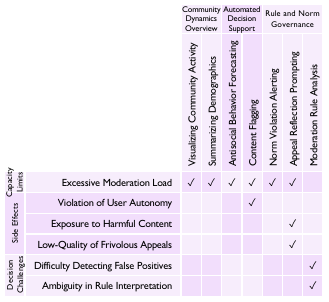}
    \caption{\yh{Design space for AI-infused system to support content moderation. }
    }
    \Description{Design space for content moderation.}
    \label{fig: design space moderation}
\end{figure*}
\Cref{fig: design space moderation} outlines the design space for content moderation, \yh{revealing two critical architectural rationales.
First, addressing the capacity limits of moderators relies on a dual-sided strategy}. On the user side, systems aim to curb inappropriate behaviors \yh{at the source through mechanisms like norm violation alerting}. 
On the moderator side, systems reduce the workload of identifying misbehaviors and handling norm violations. 
Second, \yh{most systems function as assistive technologies, leaving the final decision to the moderator to mitigate algorithmic errors and maintain user trust.
However, this design choice requires moderators} to go through the original toxic content before making a decision, which implies the tension between the pursuit of decision fairness and the challenge of exposure to harmful content.


\subsection{Human Support for AI‑Infused System Functionality}
\label{sec: human-supports}

As discussed above, AI‑infused systems achieve strong performance through awareness of human intentions and community context. However, \yh{they often require human verification or control to adapt to complex real‑world scenarios}. Examining how systems incorporate human support can reveal effective modes of human–AI collaboration and opportunities for improvement.
In this context, “human support” refers to intentional inputs that shape, tune, or enable AI functionalities, \yh{distinct from routine interactions (e.g., chatting) that merely serve as passive data sources.}
Based on our corpus, we identified four main categories of human support: 

\textbf{Feeding} involves humans actively providing data \yh{to enable system tasks}—such as contributing annotated examples for training \cite{gao2018burst, he2023cura} or tagging content to aid summarization \cite{zhang2018making}, or sharing personal social media data that the AI can use as an external knowledge source \cite{shin2021blahblahbot}.

\textbf{Calibrating} entails providing feedback or corrections to improve systems' performance or align them to humans' judgments. Examples include \yh{rectifying misclassifications} in harmful content detection \cite{hee2024brinjal, rifat2024combating, jahanbakhsh2023exploring} or stance prediction \cite{gao2018burst}, and \yh{contesting false performance assessments in group contexts} \cite{do2023inform, do2023err}.

\textbf{Customizing} \yh{allows users to tailor system behavior to individual preferences, either explicitly or implicitly.} For instance, users could select the content scope for summarization \cite{sun2016videoforest}, or \yh{revise model settings} to realize interactive topic modeling \cite{hoque2015convisit}. Humans could also use chat history to implicitly provide preferences for systems to generate personalized summarizations \cite{tepper2018collabot}.

\textbf{Configuring} refers to defining rules, parameters, or commands that guide system behavior. For example, humans could define or revise moderation rules \cite{chandrasekharan2019crossmod, lai2022human, song2023modsandbox}, set parameters for content filtering \cite{park2016supporting} or overspeaking detection \cite{bagmar2022analyzing}, and adjust weights in generation tasks \cite{zhang2024mentalimager}. Users may also issue natural language commands to assign tasks directly to the system \cite{toxtli2018understanding}.

\subsection{Cross-Theme Co-Occurrences and Design Tensions: Lessons Learned}
\label{sec: 7.1.2}

\zyh{To conclude our analysis of RQ1, we provide a deeper analytical engagement with our findings. Through cross-theme mappings of functionalities and challenges, we uncover underlying design strategies and potential design tensions}

\subsubsection{Lesson 1: Design for the Ecosystem, Not the Silo.}
\yh{
Current design paradigms often treat community challenges as isolated problems requiring targeted solutions. This is visibly reflected in \Cref{fig: RQ1 RQ2 co-occurrence}, where the highest density of research clusters along the diagonal (e.g., using \textit{Moderation Support} to address \textit{Moderation Limitations}). While logical, this ``siloed'' approach ignores the interconnected nature of community dynamics. Our analysis of the sparse but critical ``off-diagonal'' co-occurrences suggests a necessary shift in design strategy for the entire community lifecycle.

Existing successful designs illustrate how interventions in one domain can effectively mitigate challenges in another. For instance, what manifests as a contribution barrier may stem less from a lack of motivation and more from consumption bottlenecks \cite{liu2023coargue}. This suggests that effective interventions might focus less on prompting users to write, and more on reducing the burden of reading to unlock their capacity to contribute. 
Similarly, moderation challenges can be addressed through social engineering rather than ``hard'' enforcement. Systems can act as social catalysts to trigger community self-correction, offering a scalable alternative to formal oversight \cite{seering2020takes}.

Beyond existing patterns, the empty quadrants in our analysis highlight under-explored design opportunities. First, the gap connecting \textit{Contribution Support} to \textit{Moderation Limitations} points toward a strategy of pre-emptive moderation. Rather than over-investing in downstream detection, we argue that the most scalable moderation strategy is actually contribution tools: providing upstream writing assistance (e.g., tone checks) to reduce the influx of toxic content at the source.

Conversely, the gap connecting \textit{Moderation Support} to \textit{Contribution Barriers} highlights a design philosophy: treating moderation as a pedagogical instead of a punitive function. 
Current opaque ``act and hide'' tactics (e.g., silent deletion) of AI moderators may suppress participation and erode confidence.
Future systems should instead transform rejection into feedback. By explaining the rationale behind violations and suggesting specific revisions, AI turns a gatekeeping mechanism into a source of constructive guidance, thereby lowering entry barriers.
}

\vspace{\baselineskip}
\noindent 
Beyond these cross-theme mappings, our analysis also uncovers potential design tensions across themes.  
By anticipating these trade-offs, designers can make intentional choices about when, how, and to what extent to deploy specific AI-infused functionalities.

\subsubsection{Lesson 2: Motivate for Quality, Not Just Quantity}
\yh{Current strategies often prioritize the quantity of participation (e.g., post frequency) over content value. However, pushing users to contribute without ensuring they have the capacity to do so risks scaling noise rather than in-depth discourse \cite{do2023inform}.
We argue that motivation support is most effective when paired with ability support. Future designs should move beyond simple prompts that demand output; instead, incentives should target the effort to improve content---for example, rewarding users for adopting structural scaffolding. By aligning rewards with the writing process rather than just the result, AI shifts its role from a cheerleader for activity to a coach for quality.}

\subsubsection{Lesson 3: Balance Content Prioritization with Diversity}

Content prioritization, while effective in mitigating information overload, can inadvertently reduce perspective diversity by repeatedly surfacing similar viewpoints \cite{he2023cura}, potentially exacerbating filter bubbles and polarization. 
\yh{Designers should not treat prioritization as a black box. Instead, the field must move toward transparency: explicitly signaling ranking criteria helps users recognize the boundaries of their informational bubble. Furthermore, future systems should actively inject diversity (e.g., surfacing well-reasoned opposing arguments alongside popular ones). The critical direction is to broaden exposure without compromising consumption efficiency, preventing prioritization from becoming polarization.}

\subsubsection{Lesson 4: Preserve Authenticity While Enhancing Clarity}

\yh{While AI-driven information processing aids sensemaking, it may strip away the linguistic idiosyncrasies and ``messiness'' that signal the authenticity of human–human interactions \cite{zhang2025coknowledge}. Current designs risk creating a homogenized community voice where users feel they are interacting with bots rather than people. Designers should not treat raw user expression as ``noise'' to be cleaned. Instead, future systems should adopt a layered presentation strategy : preserving raw, unfiltered interactions as the immutable substrate of the community, while offering AI synthesis only as an optional overlay . This ensures that the pursuit of efficiency does not erode the sense of co-presence essential for social connection.}

\subsubsection{Lesson 5: Manage AI Output via Progressive Disclosure}
\yh{Unrestrained AI generation may overwhelm users with excessive information, increasing cognitive load and interpretation burden~\cite{piccolo2021agents}. 
In such situations, rather than presenting content as monolithic blocks, designers should adopt progressive disclosure. Systems can deliver content incrementally---by presenting a concise summary or ``hook'' first, and expanding details only upon interaction. This shift from information broadcasting to interaction-driven exploration ensures that AI respects the finite resource of community attention.}

\subsubsection{Lesson 6: Introduce Dissent with Emotional Safeguards.}

Efforts to facilitate exposure to diverse perspectives---such as implementing agents to play devil's advocates \cite{chiang2024enhancing}---can enhance deliberative quality \cite{nemeth2001devil, nemeth1983creative} but may also induce emotional strain or interaction barriers. 
\yh{Merely injecting opposing viewpoints is insufficient if it alienates the user. Thus,} designers should pair such interventions with emotional‑management functionalities, such as delivering positive social cues or pruning negative content (Table \ref{tab:codebook2}), to sustain engagement while preserving psychological safety.

\section{RQ2: How have existing studies evaluated the impact of AI-infused systems in online communities}
\label{sec: RQ3}

\begin{figure*}[htbp]
    \centering
    \includegraphics[width=1\textwidth]{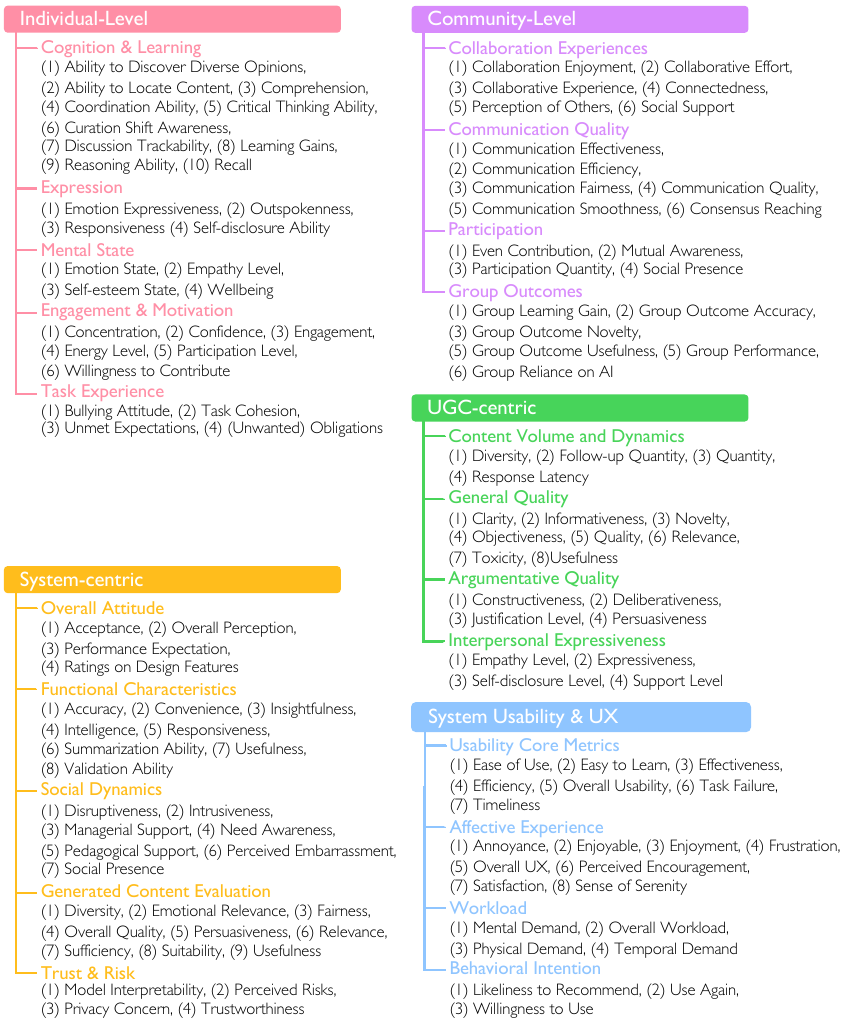}
    \caption{\yh{Evaluation metrics of the impact of AI-infused systems in online communities, categorized into five types of metrics: \textbf{Individual-level}, \textbf{Community-level}, \textbf{UGC-centric}, \textbf{System-centric}, and \textbf{System Usability \& UX}. }}
    \label{fig: RQ2 overview tree diagram 2col}
\end{figure*}

In this section, we examine how existing studies evaluated the impact of AI-infused systems in online communities. We synthesize the evaluation metrics employed across the corpus and annotate each metric with its associated data collection method. 
Based on the nature of what is being evaluated, we group the metrics into five main categories: \textbf{Individual-level, Community-level, UGC-centric, System-centric, and System Usability \& UX}. 
\yh{An overview of the codebook is presented in \Cref{fig: RQ2 overview tree diagram 2col}.}
\yh{A full breakdown of the results is included in \Cref{appendix: RQ2 evaluation metrics}.}

\normalsize

\subsection{Individual-level}

Individual-level metrics reflect each member’s own capacity and state when engaging in various behaviors under AI intervention. \textbf{Cognition and learning} metrics are mostly adopted in the evaluation of consumption support systems, assessing users' ability to absorb, navigate, and critically reflect on content. Example metrics include comprehension and recall \cite{zhang2025coknowledge}, ability of critical thinking \cite{tanprasert2024debate}, and locate relevant information \cite{hoque2015convisit}. Some studies also evaluate users’ awareness of content curation shifts \cite{he2023cura} and their coordination ability \cite{avula2022effects}.
Many contribution support studies adopt \textbf{expression} metrics, which focus on how users articulate their thoughts and emotions, such as emotion expressiveness \cite{benke2020chatbot, niksirat2023potential}, self-disclosure \cite{zhang2024mentalimager}, responsiveness \cite{hadfi2021argumentative}, and outspokenness \cite{kim2021moderator, avula2022effects}.
Several subdimensions are more widely adopted across different types of systems. \textbf{Mental state} metrics capture users’ affective and psychological states during community interaction. These include measures of emotional state \cite{wang2021cass, benke2020chatbot}, empathy \cite{zhang2024mentalimager}, self-esteem \cite{narain2020promoting}, and overall wellbeing \cite{narain2020promoting}. \textbf{Engagement and motivation} metrics relate to users’ involvement and persistence in participation.  Frequently used indicators include participation level \cite{hoque2015convisit, peng2020exploring, do2022should, zhang2018making, xia2022persua, aiyer2023taming} (e.g., how much time is spent contributing and how many times users contribute), willingness to contribute \cite{zhang2024mentalimager, shin2022chatbots}, concentration \cite{wu2024comviewer, liu2022planhelper}, and confidence (as a proxy for self-efficacy) \cite{de2024assessing, liu2023coargue, zhang2025coknowledge, wu2024comviewer, peng2020exploring, xia2022persua}.
\textbf{Task-related} metrics, which capture individual's perceptions of the collaborative process and their interpersonal experiences within task contexts, include measures such as task cohesion \cite{shin2023introbot, kim2021moderator} and the degree of perceived obligation \cite{benke2020chatbot, niksirat2023potential}.

\subsection{Community-level}

Community-level metrics capture the community dynamics and collective outcomes enabled or shaped by AI-infused systems. Most of these metrics are adopted in studies on interaction mediation systems.
\textbf{Collaboration experience} metrics are commonly used to assess how users perceive the interpersonal dynamics during group tasks. In addition to evaluating the overall experience \cite{do2023inform, do2023err}, these metrics include collaboration effort \cite{avula2019embedding}, enjoyment \cite{avula2019embedding}, perceived connectedness \cite{nordberg2019designing, shin2021blahblahbot}, and social support \cite{benke2020chatbot, niksirat2023potential}. Some studies also examine users’ perception of others \cite{avula2022effects, kumar2014triggering}.
\textbf{Communication quality} metrics focus on the process and structure of interaction. Beyond assessing overall quality  \cite{shin2021blahblahbot, kim2020bot, chiang2024enhancing, shin2023introbot, kim2021moderator, kumar2014triggering}, these metrics include communication effectiveness  \cite{kim2021moderator, kim2020bot}, efficiency \cite{kim2020bot, nordberg2019designing, kim2021moderator}, fairness \cite{kim2021moderator}, and smoothness \cite{avula2022effects}. In the context of deliberative discussions, some studies also evaluate the efficiency of consensus reaching \cite{han2023redbot, avula2022effects} to examine how well AI systems facilitate productive and civil dialogue.
\textbf{Participation} metrics measure how actively and effectively members engage in community activities. Common indicators include participant quantity  \cite{wang2021cass} and their social presence \cite{de2024assessing, benke2020chatbot, do2022should}. Even contribution  \cite{de2024assessing, kim2020bot, do2022should, do2023inform, kim2021moderator, do2023err} is another frequently used metric to assess the balance of participation dynamics. Mutual awareness—users’ awareness of others’ needs, emotions, or stances, whether at the individual or group level—is also commonly evaluated \cite{avula2019embedding, han2023redbot, avula2018searchbots, avula2022effects, gao2018burst, peng2019gremobot}.
Lastly, \textbf{group outcome} metrics assess the quality and utility of collaborative outputs. These include performance-related indicators such as accuracy \cite{chiang2024enhancing, do2022should}, learning gain \cite{do2022should}, novelty \cite{do2023inform}, and usefulness \cite{do2023inform}.

\subsection{UGC-centric}

UGC-centric metrics focus on evaluating various aspects of content generated by users, most of which are adopted in systems that support contribution or mediation. By assessing the quality and characteristics of AI-supported UGC, these metrics reflect the system’s effectiveness in achieving its intended goals.
\textbf{Content volume and dynamics} metrics encompass several widely used indicators. The most common among them is quantity \cite{ito2022agent, hadfi2021argumentative, kim2020bot, wang2021cass,he2023cura, yeo2024help, do2023inform, shin2023introbot, do2023err, toxtli2018understanding}, which not only captures the number of messages, posts, or comments generated, but also the number of unique ideas or arguments identified within them. Other frequently used metrics include diversity \cite{ito2022agent, kim2020bot, yeo2024help, kim2021moderator, kim2021starrythoughts} (e.g., of opinions) and dynamic measures such as follow-up quantity \cite{ito2022agent, wang2021cass} (e.g., the number of responses triggered by a comment), unanswered contributions \cite{wang2021cass} (e.g., the count of posts without replies), and response latency \cite{wang2021cass} (e.g., the time interval between contributions).
\textbf{General quality} metrics assess broad content attributes such as clarity \cite{do2023err}, informativeness \cite{xia2022persua, do2023err}, usefulness \cite{do2023err}, objectiveness \cite{perez2018collaborative}, novelty \cite{do2023err}, and relevance \cite{do2023err}. Toxicity \cite{atreja2024appealmod} is also frequently measured in moderation-related studies to identify harmful or inappropriate content or appeals.
In comparison, \textbf{argumentative quality} metrics are primarily used to evaluate the reasoning strength and discursive structure of content. These include justification level \cite{yeo2024help}, persuasiveness \cite{xia2022persua}, deliberativeness \cite{kim2021moderator}, and constructiveness \cite{yeo2024help}.
Finally, \textbf{interpersonal expressiveness} metrics reflect the emotional and social qualities conveyed through user content. Typical indicators include self-disclosure \cite{zhang2024mentalimager}, empathy \cite{shin2022exploring}, expressiveness \cite{yeo2024help}, and the degree of social support \cite{shin2022exploring, wu2024comviewer} embedded in responses.

\subsection{System-centric}

System-centric metrics reflect various attributes of AI-infused systems, captured through both their demonstrated performance and users’ perceptions. These metrics are widely adopted across all categories of studies in our corpus. Metrics under \textbf{overall attitude} describe users’ general impressions of the system, including acceptance \cite{do2023err, do2023inform}, performance expectations \cite{do2023inform}, and ratings of specific design features \cite{xia2022persua}. \textbf{Functional characteristics} focus on system’s technical performance and practical utility. Common indicators include accuracy \cite{da2020crokage, chandrasekharan2019crossmod, he2023cura, peng2020exploring, kim2021moderator, xia2022persua, han2023redbot, do2023err}, usefulness \cite{de2024assessing, gao2018burst, benke2020chatbot, shin2022chatbots, liu2023coargue, zhang2025coknowledge, perez2018collaborative, wu2024comviewer, hoque2015convisit, peng2024designquizzer, do2022should, wang2020jill, song2023modsandbox, liu2022planhelper, avula2018searchbots, avula2022effects, cao2023visdmk}, convenience \cite{ueda2021cyberbullying}, and responsiveness \cite{avula2022effects}. Some studies also assess the perceived intelligence \cite{do2023inform} or insightfulness \cite{cao2023visdmk} the system displays during support processes. The \textbf{social dynamics} dimension examines how the system shapes interpersonal interactions and community atmosphere. Representative metrics include social presence \cite{do2022should, de2024assessing}, pedagogical or managerial support \cite{lee2020solutionchat}, awareness of user needs  \cite{avula2022effects}, and potential downsides such as intrusiveness \cite{do2022should} or embarrassment \cite{do2023inform, do2022should}. When systems produce responses, suggestions, or summaries, \textbf{generated content evaluation} becomes essential. Relevant metrics includes the content’s overall quality \cite{da2020crokage, zhang2024mentalimager}, usefulness \cite{xu2017answerbot, da2020crokage, gao2023know, avula2022effects}, relevance \cite{xu2017answerbot, da2020crokage, gao2023know, zhang2024mentalimager}, fairness \cite{han2023redbot}, and persuasiveness \cite{govers2024ai}. In addition, \textbf{trust and risk} metrics capture users’ evaluations of system transparency, safety, and reliability—particularly important in sensitive or high-stakes contexts. These include trustworthiness \cite{lai2022human, shin2023introbot, avula2018searchbots, cao2023visdmk}, model interpretability \cite{lai2022human}, perceived risks \cite{peng2020exploring}, and privacy concerns \cite{benke2020chatbot, peng2020exploring, niksirat2023potential}.

\subsection{System Usability \& UX}

Evaluating usability and user experience is a widely recognized objective in system design and assessment. Accordingly, more than half of the papers in our corpus adopt relevant metrics, many of which are drawn from established instruments such as the System Usability Scale (SUS) \cite{lewis2018system} and NASA Task Load Index (NASA-TLX) \cite{hart2006nasa}. \textbf{Usability core metrics} include overall usability \cite{shin2022chatbots, liu2023coargue, zhang2025coknowledge, liu2022planhelper, aiyer2023taming}, effectiveness \cite{lai2022human, zhang2024see}, efficiency \cite{lai2022human, han2023redbot, aiyer2023taming, zhang2017wikum}, ease of use \cite{gao2018burst,wu2024comviewer, choi2023convex, peng2024designquizzer, avula2019embedding, shin2022exploring, xia2022persua, aiyer2023taming}, ease of learning \cite{xia2022persua}, capturing how intuitively and efficiently users can operate the system. \textbf{Affective experience} metrics reflect users’ emotional responses toward the system, such as satisfaction \cite{ito2022agent, gao2018burst, benke2020chatbot, peng2024designquizzer, shin2022exploring, han2023redbot, avula2022effects, kumar2014triggering, liu2022planhelper, piccolo2021agents}, enjoyment \cite{gao2018burst, hoque2015convisit}, frustration \cite{avula2022effects}, annoyance \cite{piccolo2021agents, peng2019gremobot}, sense of serenity \cite{liu2022planhelper}, and perceived encouragement \cite{peng2019gremobot}. \textbf{Workload} metrics are often reported as overall workload \cite{atreja2024appealmod, zhang2025coknowledge, wu2024comviewer, avula2019embedding, chiang2024enhancing, lai2022human, zhang2018making, liu2022planhelper, avula2018searchbots, aiyer2023taming, avula2022effects,zhang2017wikum} or broken down into mental, physical, and temporal demands \cite{avula2022effects, zhang2018making}. Additionally, some studies adopt \textbf{behavioral intention} metrics, which measure users’ willingness to use \cite{wu2024comviewer, ueda2021cyberbullying}, intention to use again \cite{de2024assessing, peng2024designquizzer, xia2022persua}, and likelihood to recommend the system to others \cite{niksirat2023potential}.

\section{Discussion}

\zyh{Drawing from the empirical analysis across both RQs, this section transitions from our RQ-specific insights to a prescriptive roadmap for both practice and research. Section \ref{sec: 7.2} outlines actionable design considerations, guiding practitioners to navigate complex sociotechnical dynamics. Section \ref{sec: 7.3} highlights underexplored frontiers in the current literature, proposing future research opportunities to foster more inclusive and holistic AI-infused community ecosystems. }

\subsection{Design Considerations}
\label{sec: 7.2}
\subsubsection{Transforming Workload into Engagement}

\yh{
Current design paradigms often strive for ``frictionless'' integration, viewing any additional user workload or workflow disruption as a failure to be eliminated. We argue against this zero-friction ideal. Given that hallucinations and inaccuracies remain unavoidable, the goal should be reframing inevitable extra work (e.g., verification) as a mechanism for meaningful engagement.
Specifically, instead of obscuring AI uncertainty to mimic human-like confidence, systems should introduce intentional friction—such as interfaces that explicitly invite users to audit AI outputs. This shifts users from passive recipients to active overseers, preserving agency and critical thinking. Similarly, rather than forcing users to adapt to black-box workflows, future designs should not just explain how systems work, but explicitly signal when and what human support (Section \ref{sec: human-supports}) is required. The challenge is not removing humans from the loop to save time, but designing better loops that strategically leverage human insight.

}

\subsubsection{AI-Generated Content Quality and Safety}
\yh{The proliferation of generative AI risks flooding communities with plausible but empty content that is syntactically fluent but lacks social depth or factual grounding~\cite{yeo2024help, zhang2024mentalimager}. 
Instead of replacing human communication, future designs should scaffold interaction, helping users synthesize context or structure thoughts while ensuring the communicative act remains human-driven to preserve authenticity.
Furthermore, regarding safety (especially for vulnerable groups \cite{tanprasert2024debate}), systems should align with local norms and offer interactive control, operating as transparent partners that respect user autonomy and community ecology.}

\subsubsection{Contextualized System Design}

\yh{
Current interventions often react to static content rather than adapting to user states that are dynamically changing. Future designs should focus more on continuously analyzing upstream drivers, such as psychological triggers of user behavior (e.g., identity alignment~\cite{tanprasert2024debate} or perceived autonomy \cite{rifat2024combating}), to intervene at the source.
However, we caution against unchecked contextualization. Deep sensing of user context, especially from the physical world~\cite{chen2023GapSynergyEnhancing}, risks crossing the line into surveillance~\cite{benke2020chatbot}. Future designs must balance the trade-off between contextualization and privacy, ensuring support does not feel like monitoring.
}

\subsubsection{Emotional Design}

\yh{Current designs often prioritize task efficiency, assuming that performance improvements lead to user satisfaction. However, even high-performing systems can provoke frustration~\cite{avula2022effects}, discord~\cite{chiang2024enhancing}, or make users feel clinically ``diagnosed'' rather than supported \cite{shin2022exploring}. 
Therefore, future systems should treat emotional experience as a primary metric distinct from task performance. 
Designs should aim for affective alignment---for example, by carefully selecting language styles~\cite{wang2021cass, piccolo2021agents} and managing the public-private interaction scope~\cite{do2022should}. }


\subsubsection{Navigating the Blowback Effect}
\yh{
Emerging AI strategies offer scalable, real-time capabilities to regulate complex community dynamics, yet they risk a blowback effect, where interventions exacerbate the target challenges. For instance, discussion mediation aims to mitigate polarization by injecting opposing viewpoints; however, such forced exposure may entrench polarization: users often interpret AI inputs as personal attacks or opinion manipulations, prompting them to defensively reinforce their existing beliefs rather than being open-minded~\cite{dillard2005nature,brehm1966theory}. Similarly, AI counter-misinformation tools acting as authoritative arbiters of truth may backfire by infringing on autonomy, causing users to distrust the system and reject valid corrections~\cite{brehm1966theory,piccolo2021agents}. To align with responsible AI principles, designers should employ transparent, agency-preserving strategies (such as Socratic prompting~\cite{peng2024designquizzer}) rather than opaque  verdicts to avoid activating psychological defenses.
}

\subsubsection{Ethical Concerns}
\yh{
Current AI interventions raise several ethical risks.
First, systems designed to streamline consensus or filter content risk reinforcing artificial groupthink: by prioritizing majority-aligned inputs, AI may suppress minority dissent and enforce a hollow uniformity~\cite{he2023cura}. 
Second, LLMs may amplify harmful stereotypes, as biased training data can yield skewed social depictions~\cite{tanprasert2024debate}. 
Third, over-reliance on AI can undermine the human experience. Delegating communication to agents (e.g., rewriting posts or generating summaries) risks reducing authentic interpersonal exchange to humans interacting with AI ~\cite{dyke2013enhancing}.
Thus, designs should pivot toward revealing interactional provenance by explicitly distinguishing machine logic from human expression. This transparency safeguards user agency against uniformity, exposes latent biases for scrutiny, and ensures AI supports authentic communication instead of supplanting it.
}




\subsection{Future Research Opportunities}
\label{sec: 7.3}

\subsubsection{Beyond the Current Scope: Missing Supports for Online Communities}

Although our corpus identifies a wide spectrum of challenges, \yh{existing systems neglect two foundational lifecycle stages of online communities}: \textbf{starting new communities} and \textbf{socializing newcomers}~\cite{kraut2012building}.

\yh{First, current systems favor established, well-known communities, with limited focus on newly formed ones facing the ``cold start'' problem---where insufficient content fails to attract participants, and a lack of participants hinders content generation. Designers should leverage generative AI to bootstrap activity by producing seed content (e.g., starter posts and FAQs) while clearly labeling these as artificial to maintain transparency and avoid trust issues.}

Similarly, attracting and socializing newcomers---who often lack information about the community and have low initial commitment---remains underexplored. Future researchers could use AI to generate personalized, context-aware previews of community culture, norms, and activity by matching them to individual interests and communication styles. AI can also guide onboarding through real-time, in-line norm suggestions, flagging potential missteps before posting, thereby fostering positive early experiences that reduce churn and minimize disruption.

\subsubsection{Motivating Engagement Beyond Contribution}

\yh{Our corpus reveals that while most motivational support targets contribution behavior, two critical yet overlooked forms of engagement remain under-supported: consumption and moderation.

Regarding consumption, the ICAP framework~\cite{chi2014icap} highlights that deeper learning stems from constructive  (e.g., generating explanations) and interactive (e.g., co-constructing understanding with others) engagement, yet current systems typically support only passive modes. 
Although only a few systems attempt to \textbf{transfer low-engagement consumption toward constructive or interactive forms}}, those that do often employ conversational agents to foster reflection or dialogue~\cite{peng2024designquizzer, zhang2024see}. As LLMs become more capable and accessible, such agent-based co-consumption may become a compelling paradigm for engaging users in richer, dialogic understanding. Future HCI research could explore more sophisticated strategies---such as summarization-feedback loops, curiosity-based prompting, or adaptive questioning—to facilitate deeper forms of engagement.

Another overlooked target is moderation. While communities rely on volunteer-based moderators who may begin with strong intrinsic motivation, sustaining their long-term engagement is often challenging~\cite{wang2022governing, dosono2019moderation}. Moderators frequently face emotional fatigue from repeated exposure to toxic content, and their labor is often undervalued and invisible to both the platform and the members~\cite{jhaver2019does, gillespie2018custodians}. This reveals a key design opportunity: \textbf{AI-infused systems can support moderator engagement} by making their contributions more visible, surfacing impact metrics, or providing timely positive feedback throughout the moderation process.

\subsubsection{Designing AI as Legitimate Community Members}

\yh{Most AI-infused systems function as tools augmenting individual capabilities, rather than as legitimate community members endowed with social attributes to build relationships~\cite{seering2019beyond, burke2011social}.}
While functionally oriented AI systems are undeniably valuable for addressing concrete needs, systems that assume a social role may exert broader, macro-level impacts (e.g., shaping norms, strengthening collective identity, and guiding the community’s cultural evolution or branding)~\cite{seering2019beyond}. 
\yh{This highlights a promising direction: \textbf{designing AI as legitimate community members capable of exerting social impact}.}
Based on our observations and insights from Seering et al.~\cite{seering2019beyond}, such agents may require the ability to engage in long-term, public interactions; a coherent, recognizable identity; and behavioral consistency with community norms and members’ expectations. With LLMs’ growing capabilities to maintain persona and adapt to social context~\cite{sun2024building}, this vision is increasingly within reach.

\yh{However, this paradigm entails risks.}
First, they may draw disproportionate attention to themselves, prompting members to engage with the AI rather than with one another, thereby weakening human–human interaction \cite{dyke2013enhancing}. To mitigate this, message frequency and length should be context‑sensitively moderated. For instance, if human–human interaction declines relative to human–AI interaction, the system should adopt shorter, less frequent interventions.
Second, when providing emotional support, such AIs must address ethical risks. 
\yh{In sensitive contexts,} low‑quality or inappropriate messages could cause distress, and undisclosed AI origins may lead to feelings of betrayal or disillusionment. Designers must therefore prioritize emotional safety through rigorous quality assurance---including multi‑layer review (automated, human, and community feedback), higher safety thresholds for emotion‑related outputs, and, where feasible, sentiment monitoring combined with crisis‑response protocols. Equally crucial is transparent AI identity disclosure, implemented in a clear yet integrated manner to preserve the supportive atmosphere.

\subsubsection{Advancing Inclusive Design}

\yh{Despite broader progress in AI accessibility~\cite{xu2025danmua11y, el2024ai}, our analysis reveals a critical gap: only one study in our corpus explicitly addressed users with disabilities (ADHD~\cite{nordberg2019designing}). Consequently, the field must pivot to \textbf{leveraging AI to engage people with disabilities across diverse community behaviors}. }
Different types of disabilities intersect with different community roles, each presenting distinct challenges. Thus, designers should exploit AI’s adaptability and multimodal capabilities to deliver personalized, scalable support---paving the way for more inclusive and equitable online communities~\cite{shinohara2016self, gilbert2019inclusive}.

\subsubsection{Expanding Ethical Evaluation}

\yh{Our analysis identifies a critical blind spot: the evaluation of ethical concerns (e.g., privacy and security) in AI-infused systems. 
Even when addressed, assessments are often limited to single questionnaire items~\cite{benke2020chatbot, peng2020exploring}---a superficial approach insufficient for online communities, where users frequently share sensitive personal content~\cite{shilton2018values}.
Such oversight} risks reinforcing bias, eroding user trust, and harming marginalized users~\cite{binns2018fairness, raji2019actionable}. Given the complexity of ethical considerations, future evaluations should move beyond surface-level measures and \textbf{adopt more comprehensive frameworks (e.g.,~\cite{korobenko2024towards, nikolinakos2023ethical}) to assess the ethical risks of AI-infused systems}.

\subsubsection{The Trajectory of Research}
\begin{figure*}[htbp]
    \centering
    \includegraphics[width=\textwidth]{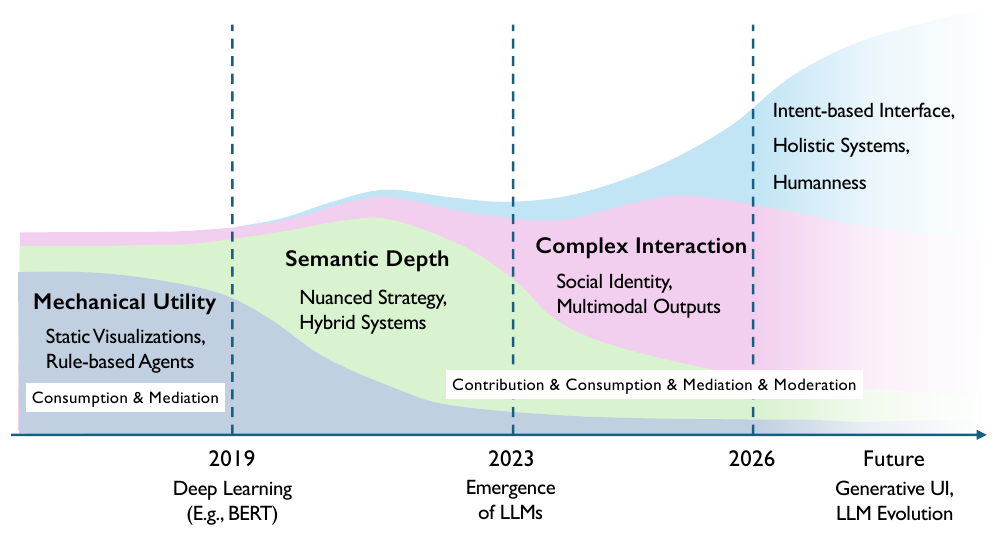}
    \caption{\yh{A streamgraph conceptualizing AI's evolution in online communities.}}
    \Description{}
    \label{fig: research trajectory}
\end{figure*}

\yh{
Synthesizing the evolution across our corpus reveals a distinct trajectory in AI's conceptualization and engineering over the past decade (\Cref{fig: research trajectory}).

Prior to 2019, research was largely defined by mechanical utility, primarily facilitating consumption through static visualizations or mediation via rule-based agents. In this era, AI agents operated with deterministic response mechanisms, where content understanding was confined to shallow metadata (e.g., post length) or basic lexical metrics (e.g., topic, sentiment). Consequently, these systems functioned as utilitarian tools with negligible social identity.

From 2019 to 2023, the proliferation of deep learning models (e.g., BERT) enabled in-depth semantic understanding of complex attributes (e.g., argument structure). Interventions diversified across all community behaviors while system functions became increasingly hybrid, allowing single systems to support multiple behaviors. Agents began exhibiting emerging social identities and nuanced interaction strategies, such as proactive posting and context-sensitive nudges.

Since 2023, the emergence of LLMs has catalyzed a significant leap in AI's capability to comprehend and generate complex conversational content, and to construct distinct social identities. While we are witnessing the nascent integration of multimodal outputs (currently primarily images), interaction mechanics (particularly turn-taking protocols) remain relatively rigid.

Looking forward, we predict a radical evolution. First, while front-end such as visualizations and standard chat interfaces have remained largely static in previous years, Generative UI \cite{leviathangenerative} promises a fundamental shift where interfaces are dynamically constructed to align with user intent. Second, more holistic systems will emerge to seamlessly integrate support across different community behaviors. Third, agents will achieve true ``humanness'' by transcending rigid protocols to make dynamic, style-congruent interaction decisions fully aligned with their personas.

}

\subsection{Limitation and Future Work}

Based on our review, we further propose some directions to continue our research.
First, our taxonomies and design insights are grounded in studies that proposed AI-infused systems, as our review intentionally scoped in system-oriented research. In future work, we aim to extend this framework by incorporating empirical studies on AI use in online communities, which may reveal complementary insights into user needs and real-world system impact. We also plan to conduct an interview study with system designers to validate and further refine the insights derived from this survey.
Second, this work establishes taxonomies of user needs and system functionalities and maps them to reveal actionable design spaces. These insights may serve as a starting point for developing configurable functionality panels within online platforms, enabling community members to personalize AI support tools based on their own challenges, thus fostering more adaptive and engaging community experiences.
Third, while our analysis focuses on challenges and functionalities within each community, future work could investigate cross-community dynamics, which may surface novel design opportunities. For instance, the early 2025 migration of TikTok users to RedNote—a Chinese online platform—triggered emergent challenges around content adaptation, cultural translation, and moderation conflicts~\cite{yuan2025love}. Studying such cross-platform transitions may offer valuable insights for designing AI-infused systems that support community integration, norm negotiation, and knowledge transfer in dynamic social environments.

\section{Conclusion}

This paper presents a systematic review of 77 studies on AI-infused systems in online communities. Our analysis offers a comprehensive overview of key aspects of system development, including the challenges these systems aim to address, the design functionalities they incorporate, and the evaluation strategies employed. Building on these taxonomies, we identify actionable design spaces, reflect on core design considerations, and outline future opportunities for leveraging AI to support community engagement. We hope this survey can assist researchers and designers in navigating design decisions and advancing the development of impactful, user-centered tools for online communities.

\bibliographystyle{ACM-Reference-Format}
\bibliography{main}

\appendix

\section{Key Challenges in Online Communities}

\begin{landscape}
\scriptsize

\renewcommand{\arraystretch}{0.8}

\label{tab:codebook1}

\begin{longtable}{p{1.8cm} p{2.5cm} p{6.5cm} p{8cm}}

\caption{\scriptsize Fundamental challenges in online communities addressed by the AI-infused systems in our corpus. The table organizes challenges into four themes: \textbf{Contribution Barriers, Consumption Obstacles, Relational Challenges, and Moderation Limitations}, each with corresponding subthemes, codes, and the definition of each code.}
\\

\toprule
\textbf{Theme} &
  \textbf{Subtheme} &
  \textbf{Code} &
  \textbf{Definition} \\* \midrule
\endfirsthead

\toprule
\textbf{Theme} &
  \textbf{Subtheme} &
  \textbf{Code} &
  \textbf{Definition} \\* \midrule

\endhead
%
\bottomrule
\endfoot
\endlastfoot
\multirow{9}{*}{\parbox{1.8cm}{Contribution \\Barriers}} &
  \multirow{3}{*}{Motivational Barriers} &
  Lack of Confidence in Contribution \cite{liu2023coargue} &
  Underestimation of the value of one's input and the belief that only highly polished or expert-level content is worth sharing. \\* \cmidrule{3-4}
 &
   &
  Anticipated Contribution Burden \cite{liu2023coargue} &
Perception that contribution to the online community is too time-consuming or cognitively or emotionally demanding.   \\* \cmidrule{3-4}
 &
   &
  Difficulty Maintaining Engagement \cite{liu2023coargue, shin2022exploring} & Difficulty in sustaining active participation in the community, particularly in fast-paced, continuous interaction environments. 
   \\* \cmidrule(l){2-4} 
 &
  \multirow{6}{*}{\parbox{2.5cm}{Cognitive and\\ Expressive Barriers}} &
  Difficulty Locating Contribution Angle \cite{liu2023coargue} & Difficulty in locating a meaningful entry point to contribute, unsure of how or where one's input can add value.
   \\* \cmidrule{3-4}
 &
   &
  Barriers to Organize Thoughts \cite{liu2023coargue} & Difficulty in organizing scattered ideas into coherent, mature responses. 
   \\* \cmidrule{3-4}
 &
   &
  Inability to Tailor Language \cite{xia2022persua} & Inability to adapt one's language style based on the topic or audience. 
   \\* \cmidrule{3-4}
 &
   &
  Limited Argumentation Skills \cite{xia2022persua} & Limited ability to construct clear and logically structured arguments and substantiate their claims with evidence.
   \\* \cmidrule{3-4}
 &
   &
  Challenges in Self-Disclosure \cite{shin2022exploring, zhang2024mentalimager} & Difficulty conveying personal experiences with sufficient clarity for support seekers in online mental health communities (OMHCs).
   \\* \cmidrule{3-4}
 &
   &
  Difficulty Providing Social Support \cite{shin2022exploring, peng2020exploring, chen2022topsbot, kumar2014triggering} &
Difficulty delivering high-quality social support for support providers.   \\* \midrule
\multirow{15}{*}{\parbox{1.8cm}{Consumption\\ Obstacles}} &
  \multirow{9}{*}{Content Flaws} &
  Overwhelming Content Volume \cite{xu2017answerbot, ito2022agent, zhang2025coknowledge, tepper2018collabot, hoque2015convisit, gao2023know, zhang2018making, zhang2024see, cao2023visdmk, zhang2017wikum} & Excessive amount of user-generated content that hinders effective navigation and engagement.
   \\* \cmidrule{3-4}
 &
   &
  Poor Content Structuring \cite{kim2020bot, gao2018burst, zhang2025coknowledge, tepper2018collabot, wu2024comviewer, peng2024designquizzer, zhang2018making, kim2021moderator, liu2022planhelper, han2023redbot, lee2020solutionchat, fu2018t, aiyer2023taming, sun2016videoforest, zhang2017wikum} & Lack of clear organization or standardized format in user-generated content. 
   \\* \cmidrule{3-4}
 &
   &
  Uneven Content Quality \cite{bagmar2022analyzing, xu2017answerbot, zhang2025coknowledge, hoque2015convisit, hoque2017cqavis, peng2024designquizzer} & Lack of consistency in the accuracy, depth, or reliability of user-generated content. 
   \\* \cmidrule{3-4}
 &
   &
  Fast-Paced Information Flow \cite{govers2024ai, lee2020solutionchat} & Excessively rapid exchange of information that overwhelms users and hinders effective processing, especially in synchronous communications. 
   \\* \cmidrule{3-4}
 &
   &
  Information Incompleteness \cite{xu2017answerbot, da2020crokage} & Lack of sufficient detail or clarity in user-generated content. 
   \\* \cmidrule{3-4}
 &
   &
  Misinformation \cite{piccolo2021agents, tanprasert2024debate, jahanbakhsh2023exploring} & Inaccurate or misleading information shared within the community. 
   \\* \cmidrule{3-4}
 &
   &
  Redundant Content \cite{zhang2025coknowledge, liu2023coargue, liu2022planhelper, zhang2017wikum} & Repetitive or duplicate user-generated content that adds little new value. 
   \\* \cmidrule{3-4}
 &
   &
  Irrelevant Content \cite{hoque2017cqavis, yeo2024help} & Off-topic or unrelated user-generated content that distracts from the main discussion.
   \\* \cmidrule{3-4}
 &
   &
  Thematic Complexity \cite{ito2022agent, zhang2025coknowledge, liu2022planhelper, zhang2024see, sun2016videoforest, cao2023visdmk} & Overly diverse or multifaceted discussion themes that complicate understanding or engagement. 
   \\* \cmidrule(l){2-4} 
 &
  \multirow{3}{*}{Retrieval Barriers} &
  Unclear Retrieval Target \cite{zhang2018making} & Uncertainty about what specific information a user is looking for during retrieval. 
   \\* \cmidrule{3-4}
 &
   &
  Query Mismatch \cite{wu2024comviewer, da2020crokage, gao2023know} & Misalignment between the user's search terms and the structure or terminology of the desired content. 
   \\* \cmidrule{3-4}
 &
   &
  Filter Bubble Effects \cite{gao2018burst, tanprasert2024debate, zhang2024see, kim2021starrythoughts} & Algorithmic personalization that constrains the content visibility and the user's exposure to diverse viewpoints. 
   \\* \cmidrule(l){2-4} 
 &
  \multirow{3}{*}{Interpretation Burdens} &
  Difficulty Maintaining Cognitive Engagement \cite{tanprasert2024debate, peng2024designquizzer, liu2022planhelper, zhang2024see} & Difficulty in staying mentally focused when processing lengthy or dense content. 
   \\* \cmidrule{3-4}
 &
   &
  Difficulty Discovering Related Content \cite{wu2024comviewer} & Difficulty in tracing connections or synthesizing information of interest across fragmented content. 
   \\* \cmidrule{3-4}
 &
   &
  Difficulty Identifying Key Information \cite{zhang2025coknowledge, wu2024comviewer, liu2022planhelper} & Difficulty in extracting valuable or representative insights from large volumes of content, especially when it lacks structure or clear signaling.
   \\* \midrule
\multirow{14}{*}{\parbox{1.8cm}{Relational\\ Challenges}} &
  \multirow{2}{*}{Emotional Dynamics} &
  Difficulty Maintaining Emotional Well-being \cite{benke2020chatbot, shin2022exploring, peng2019gremobot, narain2020promoting} & Difficulty in emotion regulation and maintaining a healthy emotional state during community interactions. 
   \\* \cmidrule{3-4}
 &
   &
  Lack of Belongingness \cite{peng2020exploring, wang2020jill} & The absence of a felt sense of being part of the community due to the loosely connected nature of the community. 
   \\* \cmidrule(l){2-4} 
 &
  \multirow{5}{*}{Interaction Barriers} &
  Difficulty Initiating Interaction \cite{shin2021blahblahbot, chen2022topsbot} & Difficulty in starting interactions due to uncertainty about appropriate topics or how to engage others.
   \\* \cmidrule{3-4}
 &
   &
  Difficulty Sustaining Interaction \cite{shin2021blahblahbot, chen2022topsbot} & Difficulty in maintaining interactions as they progress. 
   \\* \cmidrule{3-4}
 &
   &
  Unawareness of Others’ Backgrounds \cite{shin2021blahblahbot, shin2023introbot, wang2022understanding} & Lack of knowledge about community members' experiences or contexts, which impedes trust-building and weakens social cohesion among strangers.
   \\* \cmidrule{3-4}
 &
   &
  Low Social Perceptiveness \cite{benke2020chatbot, peng2019gremobot, kim2021moderator, han2023redbot, chen2022topsbot} & The limited ability to perceive collective emotional states or understand others' preferences and viewpoints, further hindering meaningful engagement. 
   \\* \cmidrule{3-4}
 &
   &
  Untimely Social Support \cite{piccolo2021agents, wang2021cass, schluger2022proactive, nordberg2019designing} & Delay in receiving emotional or informational social support when needed, usually in asynchronous communities.
   \\* \cmidrule(l){2-4} 
 &
  \multirow{2}{*}{Norm Violations} &
  Cyberbullying \cite{ueda2021cyberbullying, bagmar2022analyzing} & Repeated, intentional aggression via digital technologies targeting individuals who may struggle to defend themselves, which can cause emotional harm and suppress participation.
   \\* \cmidrule{3-4}
 &
   &
  Multi-Party Privacy Conflicts \cite{niksirat2023potential} & Sharing content involving multiple individuals without the consent of all parties. 
   \\* \cmidrule(l){2-4} 
 &
  \multirow{5}{*}{\parbox{2.5cm}{Group Coordination\\ Challenges}} &
  Uneven Participation \cite{kim2020bot, do2022should, do2023inform, kim2021moderator} & Imbalance of member contribution, where certain members under-contribute while others dominate discussions, disrupting healthy group dynamics.
   \\* \cmidrule{3-4}
 &
   &
  Task Coordination Difficulties \cite{toxtli2018understanding} & Difficulty in refining objectives, assigning roles, scheduling, and tracking progress during group collaboration, which impose a significant workload on coordinators and participants alike. 
   \\* \cmidrule{3-4}
 &
   &
  Difficulty Anticipating Group Performance \cite{gimpel2024digital} & Difficulty in predicting group collaboration outcomes, which complicates timeline estimation, resource allocation, and goal setting. 
   \\* \cmidrule{3-4}
 &
   &
  Challenges Reaching Consensus \cite{kim2020bot, shin2022chatbots, perez2018collaborative, kim2021moderator, han2023redbot} & Difficulty in reaching an agreement among community members in deliberative contexts. 
   \\* \cmidrule{3-4}
 &
   &
  Polarization \cite{govers2024ai, hadfi2021argumentative} & Deepening of opposing viewpoints in group discussions, where disagreement becomes more extreme as parties consider evidence or engage in debate, hindering constructive collaboration.
   \\* \midrule
\multirow{6}{*}{\parbox{1.8cm}{Moderation\\ Limitations}} &
  Capacity Limits &
  Excessive Moderation Load \cite{govers2024ai, atreja2024appealmod, rifat2024combating, choi2023convex, he2023cura, lai2022human, schluger2022proactive, park2016supporting, kwon2015visohc} & Overwhelming volume of user-generated content that exceeds human moderators' capacity to review and respond to inappropriate behaviors, leading to delayed interventions, overlooked violations, and reduced trust in the moderation process. 
   \\* \cmidrule(l){2-4} 
 &
  \multirow{2}{*}{Decision Challenges} &
  Difficulty Detecting False Positives \cite{song2023modsandbox} & Benign content is mistakenly flagged or removed due to inaccurate or unfair decisions in moderation, leading to user dissatisfaction and mistrust. 
   \\* \cmidrule{3-4}
 &
   &
  Ambiguity in Rule Interpretation \cite{hee2024brinjal, rifat2024combating, song2023modsandbox, chandrasekharan2019crossmod} & The lack of nuances for pre-defined moderation rules to account for context, intent, or cultural variation, and the inability for moderators to anticipate the full impact of a moderation rule before deployment. 
   \\* \cmidrule(l){2-4} 
 &
  \multirow{3}{*}{Side Effects} &
  Exposure to Harmful Content \cite{atreja2024appealmod, rifat2024combating} & Continuous exposures of moderators to large volumes of harmful or toxic content, which can negatively impact their mental health. 
   \\* \cmidrule{3-4}
 &
   &
  Low-Quality or Frivolous Appeals \cite{atreja2024appealmod} & User challenges to moderation decisions that lack substance or seriousness, increasing moderators' workload and decision fatigue. 
   \\* \cmidrule{3-4}
 &
   &
  Violation of User Autonomy \cite{jahanbakhsh2023exploring} & Moderation decisions can restrict freedom of expression or limit users’ control over what content they can access, which may not always align with users’ expectations and needs, potentially undermining trust in the system.
   \\* \bottomrule
\end{longtable}
\end{landscape}

\section{\yh{Evaluation Strategies to Assess AI-infused Systems in Online Communities}}
\label{appendix: RQ2 evaluation metrics}

\yh{We synthesized the evaluation metrics employed in the corpus
and annotated each evaluation metric with its associated data collection method.
Following a widely used taxonomy~\cite{sharp2003interaction}, we classify data collection approaches into three types: \textbf{Observation} (e.g., log analysis, behavioral traces), \textbf{Questionnaire}, and \textbf{Interview}. }



\scriptsize

\renewcommand{\arraystretch}{0.82}

\begin{longtable}[c]{|p{0.15cm}|p{2.2cm}|p{8.3cm}|p{0.25cm}?p{0.25cm}?p{0.25cm}|}

\caption{Evaluation strategies used in our corpus to assess AI-infused systems in online communities. Metrics are grouped into five categories: \textbf{Individual-level, Community-level, UGC-centric, System-centric, and System Usability \& UX}. Each metric is annotated with the data collection method(s) used: \Observation: Observation; \Questionnaire: Questionnaire; \Interview: Interview. Metrics tagged with multiple methods indicate that they have been operationalized through diverse data collection approaches in prior research.} \label{tab: codebook3} \\

\arrayrulecolor{black}\cline{2-6}
 \multicolumn{1}{c|}{} & \textbf{Subcategory} & \textbf{Metric} & \multicolumn{3}{c|}{\makecell{\textbf{Evaluation}\\\textbf{Methods}}} \\
\arrayrulecolor{black}\hline
\endfirsthead


\arrayrulecolor{black}\cline{2-6}
 \multicolumn{1}{c|}{} & \textbf{Subtheme} & \textbf{Metric} & \multicolumn{3}{c|}{\makecell{\textbf{Evaluation}\\\textbf{Methods}}} \\
\arrayrulecolor{black}\hline
\endhead

\arrayrulecolor{black}\hline
\multicolumn{6}{|r|}{\textit{Continued on next page}} \\
\arrayrulecolor{black}\hline
\endfoot

\endlastfoot

\cellcolor{applepinknormal!20} & \multirow{10}{*}{\parbox{2.2cm}{Cognition \&\\Learning}} & \cellcolor{applegrey!15} Comprehension \cite{zhang2025coknowledge}& \Observation & \Questionnaire &  \\ \arrayrulecolor{applegrey!15}\cline{4-6}
\cellcolor{applepinknormal!20} & & Recall \cite{zhang2025coknowledge} & \Observation & \Questionnaire &  \\ \arrayrulecolor{applegrey!15}\cline{4-6}
\cellcolor{applepinknormal!20} & & \cellcolor{applegrey!15} Learning Gains \cite{peng2024designquizzer, avula2019embedding}& \Observation & \Questionnaire &  \\ \arrayrulecolor{applegrey!15}\cline{4-6}
\cellcolor{applepinknormal!20} & & Critical Thinking Ability  \cite{tanprasert2024debate}&  & \Questionnaire &  \\ \arrayrulecolor{applegrey!15}\cline{4-6}
\cellcolor{applepinknormal!20} & & \cellcolor{applegrey!15} Reasoning Ability\cite{dyke2013enhancing} & \Observation &  &  \\ \arrayrulecolor{applegrey!15}\cline{4-6}
\cellcolor{applepinknormal!20} & & Ability of Discovering Diverse Opinions \cite{gao2018burst}&  & \Questionnaire &  \\ \arrayrulecolor{applegrey!15}\cline{4-6}
\cellcolor{applepinknormal!20} & & \cellcolor{applegrey!15} Ability to Locate Content \cite{hoque2015convisit}&  & \Questionnaire &  \\ \arrayrulecolor{applegrey!15}\cline{4-6}
\cellcolor{applepinknormal!20} & & Discussion Trackability \cite{lee2020solutionchat}&  & \Questionnaire &  \\ \arrayrulecolor{applegrey!15}\cline{4-6}
\cellcolor{applepinknormal!20} & & \cellcolor{applegrey!15} Curation Shift Awareness \cite{he2023cura}& \Observation &  &  \\ \arrayrulecolor{applegrey!15}\cline{4-6}
\cellcolor{applepinknormal!20} & & Coordination Ability \cite{avula2022effects}&  & \Questionnaire &  \\
        \arrayrulecolor{applegrey!40}\cline{2-6}
\cellcolor{applepinknormal!20} & \multirow{4}{*}{\parbox{2.2cm}{Expression}} & \cellcolor{applegrey!15} Emotion Expressiveness \cite{benke2020chatbot, niksirat2023potential} &  & \Questionnaire &  \\ \arrayrulecolor{applegrey!15}\cline{4-6}
\cellcolor{applepinknormal!20} & & Self-disclosure Ability \cite{zhang2024mentalimager} &  & \Questionnaire &  \\ \arrayrulecolor{applegrey!15}\cline{4-6}
\cellcolor{applepinknormal!20} & & \cellcolor{applegrey!15} Outspokenness \cite{kim2021moderator, avula2022effects}&  & \Questionnaire &  \\ \arrayrulecolor{applegrey!15}\cline{4-6}
\cellcolor{applepinknormal!20} & & Responsiveness \cite{hadfi2021argumentative}& \Observation &  &  \\
        \arrayrulecolor{applegrey!40}\cline{2-6}
\cellcolor{applepinknormal!20} & \multirow{4}{*}{\parbox{2.2cm}{Mental State}} & \cellcolor{applegrey!15} Emotion State \cite{wang2021cass, benke2020chatbot}& \Observation & \Questionnaire &  \\ \arrayrulecolor{applegrey!15}\cline{4-6}
\cellcolor{applepinknormal!20} & & Empathy Level \cite{zhang2024mentalimager}&  & \Questionnaire &  \\ \arrayrulecolor{applegrey!15}\cline{4-6}
\cellcolor{applepinknormal!20} & & \cellcolor{applegrey!15} Self-esteem State \cite{narain2020promoting}&  & \Questionnaire &  \\ \arrayrulecolor{applegrey!15}\cline{4-6}
\cellcolor{applepinknormal!20} & & Wellbeing \cite{narain2020promoting} &  & \Questionnaire &  \\
        \arrayrulecolor{applegrey!40}\cline{2-6}
\cellcolor{applepinknormal!20} & \multirow{6}{*}{\parbox{2.2cm}{Engagement \&\\Motivation}} & \cellcolor{applegrey!15} Engagement \cite{benke2020chatbot, tanprasert2024debate, peng2024designquizzer, lai2022human, niksirat2023potential, liu2022planhelper}& \Observation & \Questionnaire &  \\ \arrayrulecolor{applegrey!15}\cline{4-6}
\cellcolor{applepinknormal!20} & & Participation Level \cite{hoque2015convisit, peng2020exploring, do2022should, zhang2018making, xia2022persua, aiyer2023taming}& \Observation &  &  \\ \arrayrulecolor{applegrey!15}\cline{4-6}
\cellcolor{applepinknormal!20} & & \cellcolor{applegrey!15} Willingness to Contribute \cite{zhang2024mentalimager, shin2022chatbots} &  & \Questionnaire &  \\ \arrayrulecolor{applegrey!15}\cline{4-6}
\cellcolor{applepinknormal!20} & & Concentration \cite{wu2024comviewer, liu2022planhelper} &  & \Questionnaire &  \\ \arrayrulecolor{applegrey!15}\cline{4-6}
\cellcolor{applepinknormal!20} & & \cellcolor{applegrey!15} Confidence \cite{de2024assessing, liu2023coargue, zhang2025coknowledge, wu2024comviewer, peng2020exploring, xia2022persua}&  & \Questionnaire &  \\ \arrayrulecolor{applegrey!15}\cline{4-6}
\cellcolor{applepinknormal!20} & & Energy Level \cite{narain2020promoting} &  & \Questionnaire &  \\
        \arrayrulecolor{applegrey!40}\cline{2-6}
\cellcolor{applepinknormal!20} & \multirow{4}{*}{\parbox{2.2cm}{Task Experience}} & \cellcolor{applegrey!15} Task Cohesion \cite{shin2023introbot, kim2021moderator} &  & \Questionnaire &  \\ \arrayrulecolor{applegrey!15}\cline{4-6}
\cellcolor{applepinknormal!20} & & (Unwanted) Obligations \cite{benke2020chatbot, niksirat2023potential}&  & \Questionnaire &  \\ \arrayrulecolor{applegrey!15}\cline{4-6}
\cellcolor{applepinknormal!20} & & \cellcolor{applegrey!15} Unmet Expectations \cite{benke2020chatbot, niksirat2023potential}&  & \Questionnaire &  \\ \arrayrulecolor{applegrey!15}\cline{4-6}
\cellcolor{applepinknormal!20} \multirow{-28}{*}{\rotatebox{90}{\textbf{Individual-level}}} & & Bullying Attitude \cite{ueda2021cyberbullying}& \Observation &  &  \\
        \arrayrulecolor{black}\hline

\cellcolor{applepurplenormal!20} & \multirow{6}{*}{\parbox{2.2cm}{Collaboration\\Experiences}} & \cellcolor{applegrey!15}  Collaborative Experience \cite{do2023inform, do2023err}&  & \Questionnaire &   \\ \arrayrulecolor{applegrey!15}\cline{4-6}
\cellcolor{applepurplenormal!20} & & Collaboration Effort \cite{avula2019embedding}&  & \Questionnaire &  \\ \arrayrulecolor{applegrey!15}\cline{4-6}
\cellcolor{applepurplenormal!20} & & \cellcolor{applegrey!15}Collaboration Enjoyment \cite{avula2019embedding} &  & \Questionnaire & \\ \arrayrulecolor{applegrey!15}\cline{4-6}
\cellcolor{applepurplenormal!20} & & Social Support \cite{benke2020chatbot, niksirat2023potential}&  & \Questionnaire &  \\ \arrayrulecolor{applegrey!15}\cline{4-6}
\cellcolor{applepurplenormal!20} & & \cellcolor{applegrey!15} Connectedness \cite{nordberg2019designing, shin2021blahblahbot}&  & \Questionnaire &  \\ \arrayrulecolor{applegrey!15}\cline{4-6}
\cellcolor{applepurplenormal!20} & & Perception of Others \cite{avula2022effects, kumar2014triggering}&  & \Questionnaire &  \\
        \arrayrulecolor{applegrey!40}\cline{2-6}     
\cellcolor{applepurplenormal!20} & \multirow{7}{*}{\parbox{2.2cm}{Communication Quality}} & \cellcolor{applegrey!15}  Communication Quality \cite{shin2021blahblahbot, kim2020bot, chiang2024enhancing, shin2023introbot, kim2021moderator, kumar2014triggering}& \Observation & \Questionnaire &  \\ \arrayrulecolor{applegrey!15}\cline{4-6}
\cellcolor{applepurplenormal!20} & & Communication Efficiency \cite{kim2020bot, nordberg2019designing, kim2021moderator} &  & \Questionnaire &  \\ \arrayrulecolor{applegrey!15}\cline{4-6}
\cellcolor{applepurplenormal!20} & & \cellcolor{applegrey!15} Communication Fairness \cite{kim2021moderator}&  & \Questionnaire &  \\ \arrayrulecolor{applegrey!15}\cline{4-6}
\cellcolor{applepurplenormal!20} & & Communication Effectiveness \cite{kim2021moderator, kim2020bot} &  & \Questionnaire & \\ \arrayrulecolor{applegrey!15}\cline{4-6}
\cellcolor{applepurplenormal!20} & & \cellcolor{applegrey!15} Communication Smoothness \cite{avula2022effects}&  & \Questionnaire &  \\ \arrayrulecolor{applegrey!15}\cline{4-6}
\cellcolor{applepurplenormal!20} & & Consensus Reaching \cite{han2023redbot, avula2022effects}&  & \Questionnaire &  \\ \arrayrulecolor{applegrey!40}\cline{2-6}
\cellcolor{applepurplenormal!20} & \multirow{5}{*}{\parbox{2.2cm}{Participation}} & \cellcolor{applegrey!15} Social Presence \cite{de2024assessing, benke2020chatbot, do2022should}&  & \Questionnaire &  \\ \arrayrulecolor{applegrey!15}\cline{4-6}
\cellcolor{applepurplenormal!20} & &  Participants Quantity \cite{wang2021cass}& \Observation &  &  \\ \arrayrulecolor{applegrey!15}\cline{4-6}
\cellcolor{applepurplenormal!20} & & \cellcolor{applegrey!15} Even Contribution \cite{de2024assessing, kim2020bot, do2022should, do2023inform, kim2021moderator, do2023err}& \Observation & \Questionnaire &  \\ \arrayrulecolor{applegrey!15}\cline{4-6}
\cellcolor{applepurplenormal!20} & &  Mutual Awareness \cite{avula2019embedding, han2023redbot, avula2018searchbots, avula2022effects, gao2018burst, peng2019gremobot}&  & \Questionnaire &  \\ \arrayrulecolor{applegrey!40}\cline{2-6}
\cellcolor{applepurplenormal!20} & \multirow{6}{*}{\parbox{2.2cm}{Group Outcomes}} & \cellcolor{applegrey!15} Group Performance \cite{zhang2018making, do2023err, kumar2014triggering}& \Observation & \Questionnaire &  \\ \arrayrulecolor{applegrey!15}\cline{4-6}
\cellcolor{applepurplenormal!20} & & Group Learning Gain \cite{do2022should}& \Observation &  &  \\ \arrayrulecolor{applegrey!15}\cline{4-6}
\cellcolor{applepurplenormal!20} & & \cellcolor{applegrey!15} Group Outcome Accuracy \cite{chiang2024enhancing, do2022should} & \Observation &  &  \\ \arrayrulecolor{applegrey!15}\cline{4-6}
\cellcolor{applepurplenormal!20} & & Group Outcome Novelty \cite{do2023inform}& \Observation &  &  \\ \arrayrulecolor{applegrey!15}\cline{4-6}
\cellcolor{applepurplenormal!20} & & \cellcolor{applegrey!15} Group Outcome Usefulness \cite{do2023inform} & \Observation &  &  \\ \arrayrulecolor{applegrey!15}\cline{4-6}
\cellcolor{applepurplenormal!20} \multirow{-24}{*}{\rotatebox{90}{\textbf{Community-level}}} & & Group Reliance on AI \cite{chiang2024enhancing} & \Observation &  &  \\
        \arrayrulecolor{black}\hline

\cellcolor{applegreennormal!20} & \multirow{5}{*}{\parbox{2.2cm}{Content Volume\\and Dynamics}} & \cellcolor{applegrey!15} Quantity \cite{ito2022agent, hadfi2021argumentative, kim2020bot, wang2021cass,he2023cura, yeo2024help, do2023inform, shin2023introbot, do2023err, toxtli2018understanding} & \Observation &  &  \\ \arrayrulecolor{applegrey!15}\cline{4-6}
\cellcolor{applegreennormal!20} & & Diversity \cite{ito2022agent, kim2020bot, yeo2024help, kim2021moderator, kim2021starrythoughts} & \Observation & \Questionnaire &  \\ \arrayrulecolor{applegrey!15}\cline{4-6}
\cellcolor{applegreennormal!20} & & \cellcolor{applegrey!15} Follow-Up Quantity \cite{ito2022agent, wang2021cass}& \Observation &  &  \\ \arrayrulecolor{applegrey!15}\cline{4-6}
\cellcolor{applegreennormal!20} & & Unanswered Quantity \cite{wang2021cass}& \Observation &  &  \\ \arrayrulecolor{applegrey!15}\cline{4-6}
\cellcolor{applegreennormal!20} & & \cellcolor{applegrey!15} Response Latency \cite{wang2021cass}& \Observation &  &  \\
    \arrayrulecolor{applegrey!40}\cline{2-6}
\cellcolor{applegreennormal!20} & \multirow{5}{*}{\parbox{2.2cm}{General Quality}} & Quality \cite{kim2020bot, liu2023coargue, perez2018collaborative, wu2024comviewer, peng2020exploring, do2022should, xia2022persua, kim2021starrythoughts, do2023err}& \Observation & \Questionnaire &  \\ \arrayrulecolor{applegrey!15}\cline{4-6}
\cellcolor{applegreennormal!20} & & \cellcolor{applegrey!15} Clarity \cite{do2023err}& \Observation &  &  \\ \arrayrulecolor{applegrey!15}\cline{4-6}
\cellcolor{applegreennormal!20} & & Objectiveness \cite{perez2018collaborative}&  & \Questionnaire &  \\ \arrayrulecolor{applegrey!15}\cline{4-6}
\cellcolor{applegreennormal!20} & & \cellcolor{applegrey!15} Informativeness \cite{xia2022persua, do2023err} & \Observation & \Questionnaire &  \\ \arrayrulecolor{applegrey!15}\cline{4-6}
\cellcolor{applegreennormal!20} \multirow{-10}{*}{\rotatebox{90}{\textbf{UGC-centric}}} & & Usefulness \cite{do2023err}& \Observation &  &  \\ \arrayrulecolor{applegrey!15}\cline{4-6}
\cellcolor{applegreennormal!20} & \multirow{3}{*}{\parbox{2.2cm}{General Quality\\(Cont.)}} & \cellcolor{applegrey!15} Relevance \cite{do2023err}& \Observation &  &  \\ \arrayrulecolor{applegrey!15}\cline{4-6}
\cellcolor{applegreennormal!20} & & Novelty \cite{do2023err}& \Observation &  &  \\ \arrayrulecolor{applegrey!15}\cline{4-6}
\cellcolor{applegreennormal!20} & & \cellcolor{applegrey!15} Toxicity \cite{atreja2024appealmod}& \Observation &  &  \\
\arrayrulecolor{applegrey!40}\cline{2-6}
\cellcolor{applegreennormal!20} & \multirow{4}{*}{\parbox{2.2cm}{Argumentative Quality}} & Justification Level \cite{yeo2024help}& \Observation &  &  \\ \arrayrulecolor{applegrey!15}\cline{4-6}
\cellcolor{applegreennormal!20} & & \cellcolor{applegrey!15} Persuasiveness \cite{xia2022persua}& \Observation &  &  \\ \arrayrulecolor{applegrey!15}\cline{4-6}
\cellcolor{applegreennormal!20} & & Deliberativeness \cite{kim2021moderator} &  & \Questionnaire &  \\ \arrayrulecolor{applegrey!15}\cline{4-6}
\cellcolor{applegreennormal!20} & & \cellcolor{applegrey!15} Constructiveness \cite{yeo2024help}& \Observation &  &  \\
\arrayrulecolor{applegrey!40}\cline{2-6}
\cellcolor{applegreennormal!20} & \multirow{4}{*}{\parbox{2.2cm}{Interpersonal Expressiveness}} & Expressiveness \cite{yeo2024help} & \Observation &  &  \\ \arrayrulecolor{applegrey!15}\cline{4-6}
\cellcolor{applegreennormal!20} & & \cellcolor{applegrey!15} Self-disclosure Level \cite{zhang2024mentalimager}&  & \Questionnaire &  \\ \arrayrulecolor{applegrey!15}\cline{4-6}
\cellcolor{applegreennormal!20} & & Empathy Level \cite{shin2022exploring}&  & \Questionnaire &  \\ \arrayrulecolor{applegrey!15}\cline{4-6}
\cellcolor{applegreennormal!20} \multirow{-11}{*}{\rotatebox{90}{\textbf{UGC-centric (Cont.)}}} & & \cellcolor{applegrey!15} Support Level \cite{shin2022exploring, wu2024comviewer}& \Observation & \Questionnaire &  \\
\arrayrulecolor{black}\hline

\cellcolor{appleyellownormal!20} & \multirow{5}{*}{\parbox{2.2cm}{Overall Attitude}} & Overall Perception \cite{murgia2016among, shin2021blahblahbot, kim2020bot, benke2020chatbot, shin2022chatbots, hoque2015convisit, hoque2017cqavis, tanprasert2024debate, nordberg2019designing, jahanbakhsh2023exploring, zhang2024mentalimager, zhang2024see, kumar2014triggering}& \Observation & \Questionnaire & \Interview \\ \arrayrulecolor{applegrey!15}\cline{4-6}
\cellcolor{appleyellownormal!20} & & \cellcolor{applegrey!15} Acceptance \cite{do2023err, do2023inform}&  & \Questionnaire &  \\ \arrayrulecolor{applegrey!15}\cline{4-6}
\cellcolor{appleyellownormal!20} & &  Performance Expectation \cite{do2023inform} &  & \Questionnaire &  \\ \arrayrulecolor{applegrey!15}\cline{4-6}
\cellcolor{appleyellownormal!20} & & \cellcolor{applegrey!15} Ratings on design features \cite{xia2022persua}&  & \Questionnaire &  \\
\arrayrulecolor{applegrey!40}\cline{2-6}
\cellcolor{appleyellownormal!20} & \multirow{9}{*}{\parbox{2.2cm}{Functional\\Characteristics}} &  Accuracy \cite{da2020crokage, chandrasekharan2019crossmod, he2023cura, peng2020exploring, kim2021moderator, xia2022persua, han2023redbot, do2023err}& \Observation & \Questionnaire &  \\ \arrayrulecolor{applegrey!15}\cline{4-6}
\cellcolor{appleyellownormal!20} & & \cellcolor{applegrey!15} Usefulness \cite{de2024assessing, gao2018burst, benke2020chatbot, shin2022chatbots, liu2023coargue, zhang2025coknowledge, perez2018collaborative, wu2024comviewer, hoque2015convisit, peng2024designquizzer, do2022should, wang2020jill, song2023modsandbox, liu2022planhelper, avula2018searchbots, avula2022effects, cao2023visdmk}& \Observation & \Questionnaire & \Interview \\ \arrayrulecolor{applegrey!15}\cline{4-6}
\cellcolor{appleyellownormal!20} & &  Convenience \cite{ueda2021cyberbullying}&  & \Questionnaire &  \\ \arrayrulecolor{applegrey!15}\cline{4-6}
\cellcolor{appleyellownormal!20} & & \cellcolor{applegrey!15} Responsiveness \cite{avula2022effects} &  & \Questionnaire &  \\ \arrayrulecolor{applegrey!15}\cline{4-6}
\cellcolor{appleyellownormal!20} & &  Insightfulness \cite{cao2023visdmk}&  & \Questionnaire &  \\ \arrayrulecolor{applegrey!15}\cline{4-6}
\cellcolor{appleyellownormal!20} & & \cellcolor{applegrey!15} Intelligence \cite{do2023inform} &  & \Questionnaire &  \\ \arrayrulecolor{applegrey!15}\cline{4-6}
\cellcolor{appleyellownormal!20} & & Summarization Ability \cite{cao2023visdmk}&  & \Questionnaire &  \\ \arrayrulecolor{applegrey!15}\cline{4-6}
\cellcolor{appleyellownormal!20} & & \cellcolor{applegrey!15} Validation Ability \cite{avula2022effects}&  & \Questionnaire &  \\
\arrayrulecolor{applegrey!40}\cline{2-6}
\cellcolor{appleyellownormal!20} & \multirow{7}{*}{\parbox{2.2cm}{Social Dynamics}} & Social Presence \cite{do2022should, de2024assessing}&  & \Questionnaire &  \\ \arrayrulecolor{applegrey!15}\cline{4-6}
\cellcolor{appleyellownormal!20} & & \cellcolor{applegrey!15} Pedagogical Support \cite{lee2020solutionchat} &  & \Questionnaire &  \\ \arrayrulecolor{applegrey!15}\cline{4-6}
\cellcolor{appleyellownormal!20} & & Managerial Support \cite{lee2020solutionchat} &  & \Questionnaire &  \\ \arrayrulecolor{applegrey!15}\cline{4-6}
\cellcolor{appleyellownormal!20} & & \cellcolor{applegrey!15} Need Awareness \cite{avula2022effects}&  & \Questionnaire &  \\ \arrayrulecolor{applegrey!15}\cline{4-6}
\cellcolor{appleyellownormal!20} & & Intrusiveness \cite{do2022should}&  & \Questionnaire &  \\ \arrayrulecolor{applegrey!15}\cline{4-6}
\cellcolor{appleyellownormal!20} & & \cellcolor{applegrey!15} Disruptiveness \cite{avula2022effects} &  & \Questionnaire &  \\ \arrayrulecolor{applegrey!15}\cline{4-6}
\cellcolor{appleyellownormal!20} & & Perceived Embarrassment \cite{do2023inform, do2022should}&  & \Questionnaire &  \\
\arrayrulecolor{applegrey!40}\cline{2-6}
\cellcolor{appleyellownormal!20} & \multirow{9}{*}{\parbox{2.2cm}{Generated Content Evaluation}} & \cellcolor{applegrey!15} Overall Quality \cite{da2020crokage, zhang2024mentalimager}&  & \Questionnaire & \Interview \\ \arrayrulecolor{applegrey!15}\cline{4-6}
\cellcolor{appleyellownormal!20} & & Usefulness \cite{xu2017answerbot, da2020crokage, gao2023know, avula2022effects}  & \Observation & \Questionnaire & \Interview \\ \arrayrulecolor{applegrey!15}\cline{4-6}
\cellcolor{appleyellownormal!20} & & \cellcolor{applegrey!15} Relevance \cite{xu2017answerbot, da2020crokage, gao2023know, zhang2024mentalimager}& \Observation & \Questionnaire & \Interview \\ \arrayrulecolor{applegrey!15}\cline{4-6}
\cellcolor{appleyellownormal!20} & & Emotional Relevance \cite{zhang2024mentalimager} &  & \Questionnaire &  \\ \arrayrulecolor{applegrey!15}\cline{4-6}
\cellcolor{appleyellownormal!20} & & \cellcolor{applegrey!15} Fairness \cite{han2023redbot}&  & \Questionnaire &  \\ \arrayrulecolor{applegrey!15}\cline{4-6}
\cellcolor{appleyellownormal!20} & & Persuasiveness \cite{govers2024ai}&  & \Questionnaire &  \\ \arrayrulecolor{applegrey!15}\cline{4-6}
\cellcolor{appleyellownormal!20} & & \cellcolor{applegrey!15} Sufficiency \cite{avula2022effects}&  & \Questionnaire &  \\ \arrayrulecolor{applegrey!15}\cline{4-6}
\cellcolor{appleyellownormal!20} & & Suitability \cite{han2023redbot}&  & \Questionnaire &  \\ \arrayrulecolor{applegrey!15}\cline{4-6}
\cellcolor{appleyellownormal!20} & & \cellcolor{applegrey!15} Diversity \cite{xu2017answerbot} & \Observation &  &  \\
\arrayrulecolor{applegrey!40}\cline{2-6}
\cellcolor{appleyellownormal!20} & \multirow{4}{*}{\parbox{2.2cm}{Trust \& Risk}} & Trustworthiness \cite{lai2022human, shin2023introbot, avula2018searchbots, cao2023visdmk}&  & \Questionnaire &  \\ \arrayrulecolor{applegrey!15}\cline{4-6}
\cellcolor{appleyellownormal!20} & & \cellcolor{applegrey!15} Model Interpretability \cite{lai2022human}&  & \Questionnaire &  \\ \arrayrulecolor{applegrey!15}\cline{4-6}
\cellcolor{appleyellownormal!20} & & Perceived Risks \cite{peng2020exploring}&  & \Questionnaire &  \\ \arrayrulecolor{applegrey!15}\cline{4-6}
\cellcolor{appleyellownormal!20} \multirow{-34}{*}{\rotatebox{90}{\textbf{System-centric}}} & & \cellcolor{applegrey!15} Privacy Concern \cite{benke2020chatbot, peng2020exploring, niksirat2023potential}&  & \Questionnaire &  \\
\arrayrulecolor{black}\hline

\cellcolor{applebluenormal!20} & \multirow{7}{*}{\parbox{2.2cm}{Usability\\Core Metrics}} & Overall Usability \cite{shin2022chatbots, liu2023coargue, zhang2025coknowledge, liu2022planhelper, aiyer2023taming} &  & \Questionnaire &  \\ \arrayrulecolor{applegrey!15}\cline{4-6}
\cellcolor{applebluenormal!20} & & \cellcolor{applegrey!15} Effectiveness \cite{lai2022human, zhang2024see} & \Observation &  & \Interview \\ \arrayrulecolor{applegrey!15}\cline{4-6}
\cellcolor{applebluenormal!20} & & Efficiency \cite{lai2022human, han2023redbot, aiyer2023taming, zhang2017wikum} & \Observation & \Questionnaire &  \\ \arrayrulecolor{applegrey!15}\cline{4-6}
\cellcolor{applebluenormal!20} & & \cellcolor{applegrey!15} Ease of Use \cite{gao2018burst,wu2024comviewer, choi2023convex, peng2024designquizzer, avula2019embedding, shin2022exploring, xia2022persua, aiyer2023taming} &  & \Questionnaire &  \\ \arrayrulecolor{applegrey!15}\cline{4-6}
\cellcolor{applebluenormal!20} & & Easy to Learn \cite{xia2022persua}&  & \Questionnaire &  \\ \arrayrulecolor{applegrey!15}\cline{4-6}
\cellcolor{applebluenormal!20} & & \cellcolor{applegrey!15} Task Failure \cite{avula2022effects} &  & \Questionnaire &  \\ \arrayrulecolor{applegrey!15}\cline{4-6}
\cellcolor{applebluenormal!20} & & Timeliness \cite{liu2022planhelper} &  & \Questionnaire &  \\
\arrayrulecolor{applegrey!40}\cline{2-6}
\cellcolor{applebluenormal!20} & \multirow{8}{*}{\parbox{2.2cm}{Affective\\Experience}} & \cellcolor{applegrey!15} Overall UX \cite{shin2021blahblahbot, kim2020bot, benke2020chatbot, da2020crokage, chiang2024enhancing, zhang2018making, kim2021starrythoughts, zhang2017wikum} & \Observation & \Questionnaire & \Interview \\ \arrayrulecolor{applegrey!15}\cline{4-6}
\cellcolor{applebluenormal!20} & & Enjoyable \cite{gao2018burst, hoque2015convisit} &  & \Questionnaire &  \\ \arrayrulecolor{applegrey!15}\cline{4-6}
\cellcolor{applebluenormal!20} & & \cellcolor{applegrey!15} Frustration \cite{avula2022effects} &  & \Questionnaire &  \\ \arrayrulecolor{applegrey!15}\cline{4-6}
\cellcolor{applebluenormal!20} & & Satisfaction \cite{ito2022agent, gao2018burst, benke2020chatbot, peng2024designquizzer, shin2022exploring, han2023redbot, avula2022effects, kumar2014triggering, liu2022planhelper, piccolo2021agents} &  & \Questionnaire & \Interview \\ \arrayrulecolor{applegrey!15}\cline{4-6}
\cellcolor{applebluenormal!20} & & \cellcolor{applegrey!15} Annoyance \cite{piccolo2021agents, peng2019gremobot} &  & \Questionnaire &  \\ \arrayrulecolor{applegrey!15}\cline{4-6}
\cellcolor{applebluenormal!20} & & Enjoyment \cite{xia2022persua, avula2018searchbots, avula2022effects} &  & \Questionnaire &  \\ \arrayrulecolor{applegrey!15}\cline{4-6}
\cellcolor{applebluenormal!20} & & \cellcolor{applegrey!15} Sense of Serenity \cite{liu2022planhelper}&  & \Questionnaire &  \\ \arrayrulecolor{applegrey!15}\cline{4-6}
\cellcolor{applebluenormal!20} & & Perceived Encouragement \cite{peng2019gremobot} &  & \Questionnaire &  \\
\arrayrulecolor{applegrey!40}\cline{2-6}
\cellcolor{applebluenormal!20} & \multirow{4}{*}{\parbox{2.2cm}{Workload}} & \cellcolor{applegrey!15} Overall Workload \cite{atreja2024appealmod, zhang2025coknowledge, wu2024comviewer, avula2019embedding, chiang2024enhancing, lai2022human, zhang2018making, liu2022planhelper, avula2018searchbots, aiyer2023taming, avula2022effects,zhang2017wikum} & \Observation & \Questionnaire &  \\ \arrayrulecolor{applegrey!15}\cline{4-6}
\cellcolor{applebluenormal!20} & & Mental Demand \cite{avula2022effects}&  & \Questionnaire &  \\ \arrayrulecolor{applegrey!15}\cline{4-6}
\cellcolor{applebluenormal!20} & & \cellcolor{applegrey!15} Physical Demand \cite{avula2022effects}&  & \Questionnaire &  \\ \arrayrulecolor{applegrey!15}\cline{4-6}
\cellcolor{applebluenormal!20} & & Temporal Demand \cite{avula2022effects, zhang2018making}&  & \Questionnaire &  \\
\arrayrulecolor{applegrey!40}\cline{2-6}
\cellcolor{applebluenormal!20} & \multirow{3}{*}{\parbox{2.2cm}{Behavioral\\Intention}} & \cellcolor{applegrey!15} Willingness to Use \cite{wu2024comviewer, ueda2021cyberbullying}&  & \Questionnaire &  \\ \arrayrulecolor{applegrey!15}\cline{4-6}
\cellcolor{applebluenormal!20} & & Use Again \cite{de2024assessing, peng2024designquizzer, xia2022persua}&  & \Questionnaire &  \\ \arrayrulecolor{applegrey!15}\cline{4-6}
\cellcolor{applebluenormal!20} \multirow{-22}{*}{\rotatebox{90}{\textbf{System Usability \& UX}}} & & \cellcolor{applegrey!15} Likeliness to Recommend \cite{niksirat2023potential} &  & \Questionnaire &  \\
\arrayrulecolor{black}\hline


\end{longtable}

\end{document}